\DocumentMetadata{}
\documentclass[twoside,twocolumn,english,aps,showpacs,prl,floatfix,longbibliography]{revtex4-2}

\usepackage{array}[=2016-10-06]
\usepackage{dcolumn}
\usepackage{bm}
\usepackage{physics}
\usepackage{amsmath,graphicx}
\usepackage{mhchem}
\usepackage{tabularx}
\usepackage[dvipsnames]{xcolor}

\usepackage{pdfpages} 
\usepackage{pgffor} 

\makeatletter
\AtBeginDocument{\let\LS@rot\@undefined}
\makeatother

\pdfximage{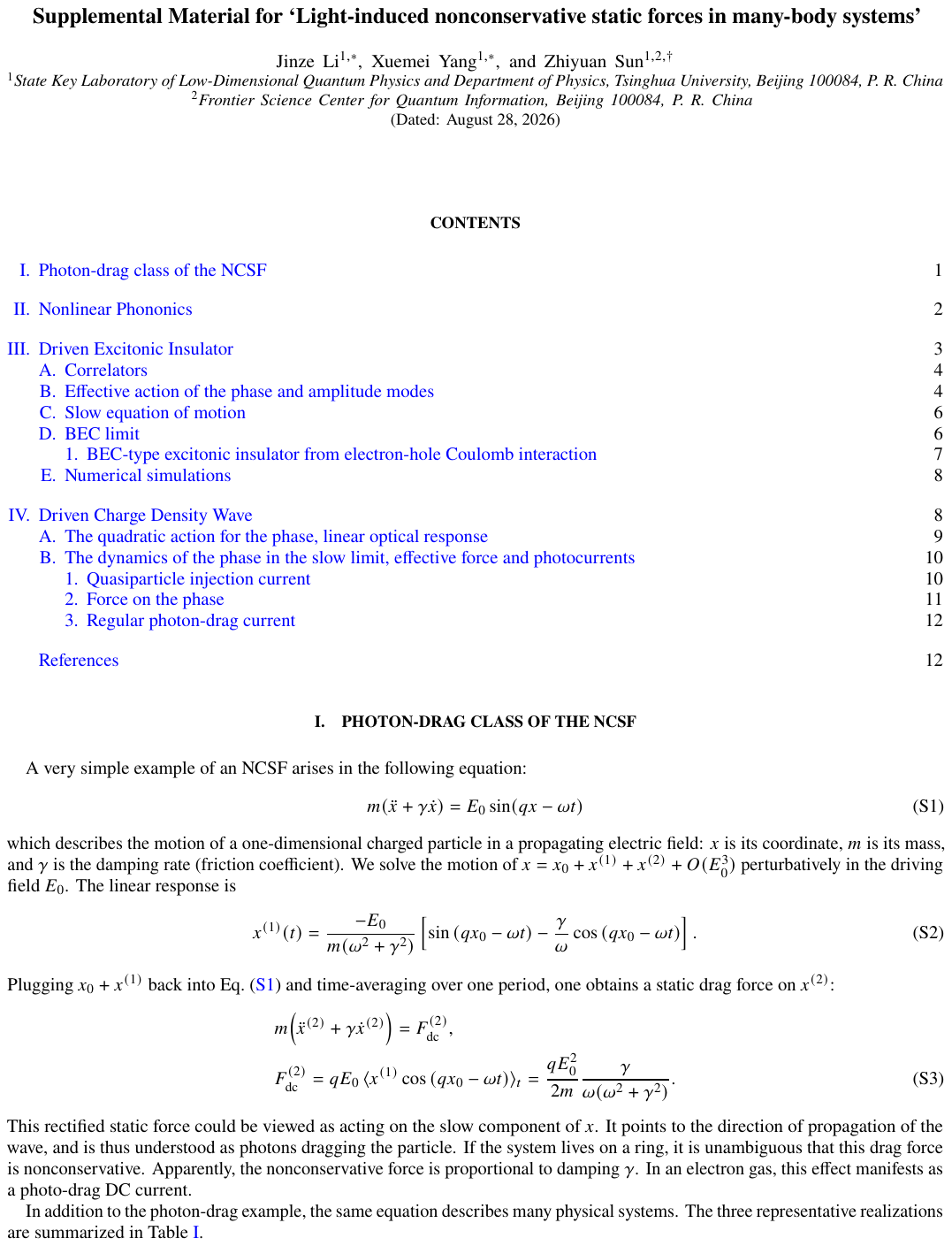}
\def\numbersupplementpages{\the\pdflastximagepages}

\newif\ifarXiv
\arXivtrue 

\usepackage[colorlinks,linkcolor=blue,anchorcolor=blue,citecolor=blue]{hyperref}
\usepackage{cleveref}
\usepackage{newpax}
\newpaxsetup{addannots=false}
\hypersetup{%
	pdftitle={NCSF}, %
	pdfauthor={Zhiyuan Sun},%
	pdfpagemode={UseNone},%
	pdfstartview={FitH},%
	citecolor=blue,%
	colorlinks=true,%
	linkcolor=blue,%
	urlcolor=blue
}

\usepackage{newtxtext}
\usepackage[varg]{newtxmath}
\usepackage{microtype}
\graphicspath{{figures/}}

\definecolor{darkred}{RGB}{139,0,0}

\newcommand{\unit}[1]{\,\mathrm{#1}} 
\newcommand{\nline}{\notag \\} 
\newcommand{\equa}[1]{Eq.~\eqref{#1}} 
\newcommand{\fig}[1]{Fig.~\ref{#1}}

\newcommand{\rom}[1]{\uppercase\expandafter{\romannumeral #1\relax}}
\renewcommand{\vec}{\mathbf}

\begin{document}

\title{Light-induced nonconservative static forces  in many-body systems}%

\author{Jinze Li$^{1,\ast}$}
\author{Xuemei Yang$^{1,\ast}$}
\author{Zhiyuan Sun$^{1,2,\dagger}$}

\affiliation{$^{1}$State Key Laboratory of Low-Dimensional Quantum Physics and Department of Physics, Tsinghua University, Beijing 100084, P. R. China\\
	$^{2}$Frontier Science Center for Quantum Information, Beijing 100084, P. R. China
}

\date{\today}


\begin{abstract}
In a quantum many-body system, a periodic drive can often generate effective static forces on the slow collective degrees of freedom.
We study the static forces generated by light on electronic order parameters in solid-state systems.
We show that the forces can have nonconservative components that originate from dissipation, i.e., optical absorption.
This effect is demonstrated in two nontrivial examples.
In excitonic insulators, via interband excitations, the light field generates a nonconservative force on the phase of the excitonic order parameter.
This force leads to an acceleration  of the phase, which manifests as a shift of the photon emission peak from an exciton condensate.
In materials with an incommensurate charge density wave, a propagating light field generates a nonconservative force on the phase of its order parameter.
It drives the charge density wave into sliding motion, leading to  a DC electric current via topological Thouless pumping.
\end{abstract}

\maketitle


\emph{Introduction---}It is well known that in systems driven by oscillating external fields, the driving field generates effective static forces on the low energy (in other words, `slow') degrees of freedom.
Simple examples include the gradient force exerted by inhomogeneous electromagnetic (EM) waves on charged particles in plasma physics~\cite{GaponovMiller1958,Aliev:1992aa}, the optical force behind optical tweezers~\cite{ashkin1970Phys.Rev.Lett.accelerationtrapping, Moffitt.2008}, and the force that stabilizes the Kapitza pendulum~\cite{Kapitza.1951}.
The conservative part of this effective force, called the `ponderomotive force'~\cite{GaponovMiller1958,landau2013electrodynamics,  Aliev:1992aa}, has been widely applied to the study of periodically driven systems~\cite{braginsky1997optical, Grimm:2000_optical_trap, buonanno2002signal,landau2013electrodynamics,wan2017control,Sun.2018, Wolff.2019_P_force_graphene, wan2018nonequilibrium,zhou2021terahertz, zhou2023vibrational, Sun.2018,
Rikhter.2024}
and recently formulated  for 
quantum many-body systems~\cite{sun2024floquet}.
It is the negative gradient of the ponderomotive potential $V_{\text{P}}[\bm{\phi}]$ that modifies the free energy landscape of the slow degrees of freedom $\bm{\phi}$ in the low-energy effective field theory. 
This landscape serves as a starting point for predicting nonequilibrium steady states of driven  many-body systems~\cite{sun2024floquet,Jiang2024, huang2025universalphasetransitionsmatter,
Diehl.2025_Josephson,
Zhou2025, hu2025microscopic, pimlott2025arxiv, erasosolarte2026, fiore2026arxiv}.

\begin{figure}
  \includegraphics[width=\linewidth]{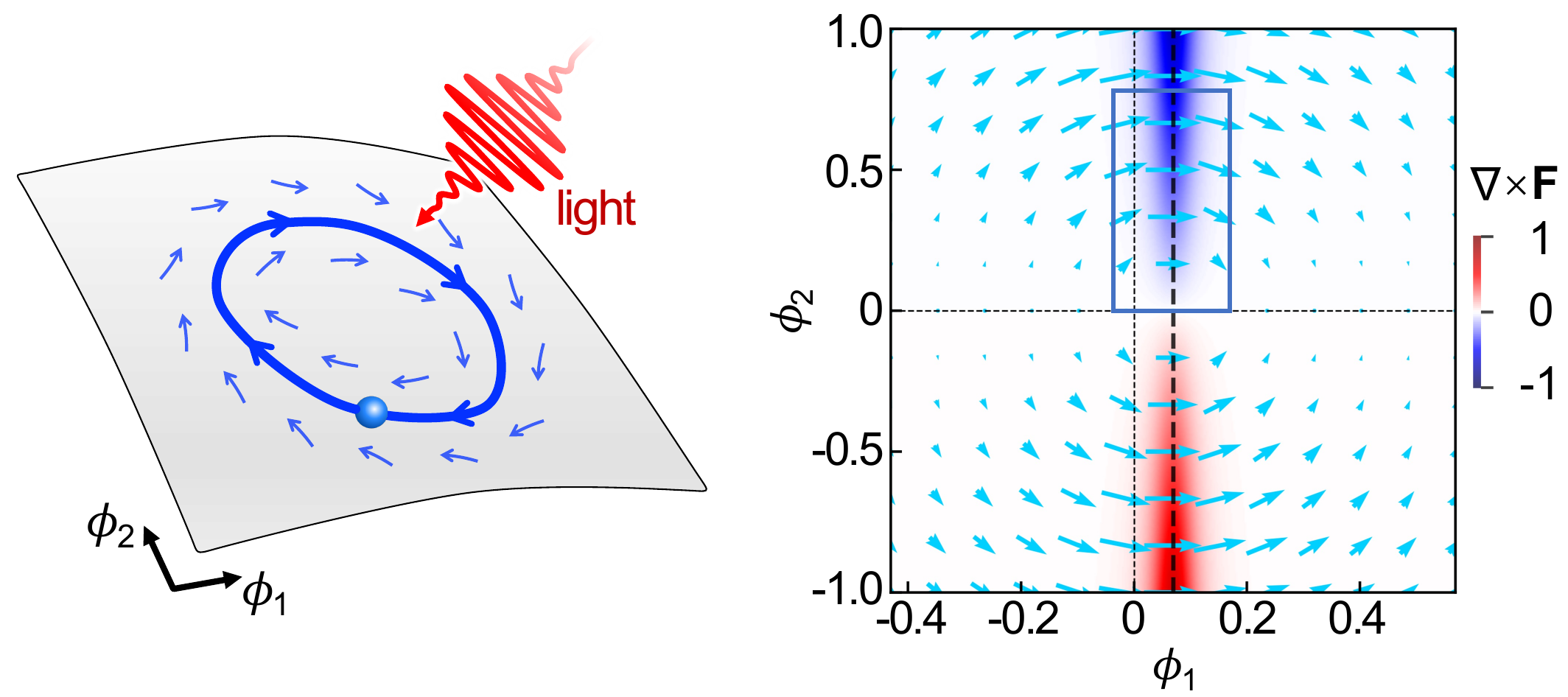}
  \caption{Left: Schematic of  light-induced nonconservative force field (blue arrows) on the manifold of slow degrees of freedom. The blue trajectory denotes a limit cycle.
  Right: Light-induced effective static force field on the slow  Raman modes $\phi_1$ and $\phi_2$ in a nonlinear phononic system. The arrows indicate the local direction and magnitude of the force while the color map shows its local curl. 
  The bold dashed line indicates the resonance configuration 
  $\omega_{\rm{IR}}^2(1 - \phi_1) = \omega^2$. 
  }
  \label{fig:prmno3}
\end{figure}

Interestingly, the effective force could have nonconservative parts in certain driven systems, such as a particle dragged by a wave or in an optical trap~\cite{Ashkin1983OpticalEarnshaw, Berry2013PhysicalCurlForces, Berry2016CurlForceDynamics}.
If the slow degree of freedom moves adiabatically on a closed path, it absorbs net energy from this static force, as shown schematically by \fig{fig:prmno3} (left).
Therefore, the time-independent low-energy effective theory exhibits intrinsically nonequilibrium characters, including limit cycles in suitable cases~\cite{Cross_Hohenberg.1993,Florian.2014, shmakov2025}.
This raises the question of whether the periodic drive can generate such forces on collective degrees of freedom in quantum matter, such as electronic order parameters.

In condensed matter physics, the most practical and vigorously studied driven systems are solid-state materials driven by light~\cite{Basov:2017_review,Kennes.2021_rmp,Zhou.2022_review,Murakami.2023_rmp_photoinduced}.
In this work, we show that light can generate nonconservative static forces (NCSFs) on the slow collective degrees of freedom in solid-state systems.
The light-induced net static force (formally a 1-form field) is
\begin{align}\label{eqn:NCSF}
\vb{F}=-\nabla V_{\rm P}+\vb{F}_{\rm NCSF},
\quad
\oint_C  \vb{F}_{\rm NCSF} \cdot  d\bm{\phi} \neq 0
\,,
\end{align}
on the manifold parametrized by $\bm{\phi}$ (e.g., electronic order parameters) where $C$ is a closed loop, see \fig{fig:prmno3} (left).
First of all, we emphasize that the NCSF ($\vb{F}_{\rm NCSF}$ in \equa{eqn:NCSF}) is rooted in \emph{dissipation}, 
which can arise either from resonant excitation of the system by the drive, or from bath-induced level broadening in an open system.
Without dissipation, Ref.~\cite{sun2024floquet} showed that the drive-induced static force is the gradient of the ponderomotive potential, which is therefore conservative. 
In the case of solid-state systems driven by light, the NCSF is thus rooted in optical dissipation.
Appendix A contains a simple understanding in the classical limit, and an explicit connection between NCSF and the dissipative parts of equilibrium response functions in a class of driven systems.

We first showcase light-induced NCSF on lattice displacements in a nonlinear phononic system.
We then demonstrate the main result in the platform of excitonic insulators: above-gap optical excitation  generates a NCSF on the phase of the excitonic order parameter, leading to persistent phase winding after a light pulse.
In bright exciton condensates that emit coherent light~\cite{Wang2019HighTemperatureExcitonCondensation,Sun2024exciton}, this NCSF leads to a shift of the emitted photons, which is 
a direct experimental observable.
Finally, we show that in incommensurate charge-density-wave (CDW) systems from Fermi surface nesting, propagating light generates a NCSF on the phase of the order parameter.
It leads to  sliding motion of the CDW, a limit-cycle state associated with a DC electric current.

\emph{Nonlinear phononic systems}~\cite{forst2011, mankowsky2014, muradudin1970, pimlott2025arxiv, erasosolarte2026}
offer a particularly transparent setting to showcase the light-induced NCSFs. A minimal description is captured by the Lagrangian following the sign convention of Ref.~\cite{sun2024floquet}:
\begin{equation}
L = \frac{1}{2} \qty[-\dot{X}^2 + \omega_{\rm{IR}}^2(1 - \phi_1) X^2] - E(t) \phi_2X + L_s[\phi_1, \phi_2],
\end{equation}
where $X$ denotes the displacement of a fast infrared (IR)-active phonon driven linearly by the electric field $E(t)=E_0\cos\omega t$. 
Two Raman modes, $\phi_1$ and $\phi_2$, couple to the IR phonon by modulating its eigenfrequency and Born
effective charge~\cite{born1996}, respectively. 
Their intrinsically slow dynamics is contained in $L_s$.
Averaging over the rapid classical oscillations of the driven IR mode yields light-induced static forces on the slow coordinates~\cite{sun2024floquet, huang2025universalphasetransitionsmatter}:
$(F_1, F_2)=-(\langle \partial_{\phi_1}(L-L_s) \rangle, \langle \partial_{\phi_2}(L-L_s) \rangle)=(\omega_{\rm{IR}}^2\langle X^2\rangle/2, \langle E(t)X(t)\rangle)$.
It reads
\begin{equation}\label{eqn:phonon_NCSF}
    \mqty(F_1 \\ F_2) =\frac{ \abs{E_0}^2 \omega_{\rm{IR}}^2 \phi_2/2}{\left[\omega_{\rm{IR}}^2(1-\phi_1) - \omega^2 \right]^2 + \gamma^2 \omega^2}
    \mqty(\phi_2/2 
    \\ 
    1-\phi_1-\frac{\omega^2}{\omega_{\rm{IR}}^2})
\end{equation}
where a damping rate $\gamma$ for the IR mode has been added.
Because of damping, this force field has nonzero curl in the $(\phi_1,\phi_2)$ plane, as illustrated in Fig.~\ref{fig:prmno3} (right).
The loop integral of the force is obviously nonzero around the blue rectangle there.
A concrete realization could be achieved in perovskite oxides with soft Raman phonons such as \ce{PrMnO3}~\cite{subedi2014}, where significant anharmonic couplings between IR- and Raman-active phonons are known to be present.
The forces in \equa{eqn:phonon_NCSF} will modify the stable location of the slow Raman coordinates, or even support limit cycles in suitable parameter regimes. 
It corresponds to slow  modulations of lattice distortion and other observables, such as the optical conductivity~\cite{rini2007}, which could be measured in pump-probe experiments.


\begin{figure}
  \includegraphics[width=\linewidth]{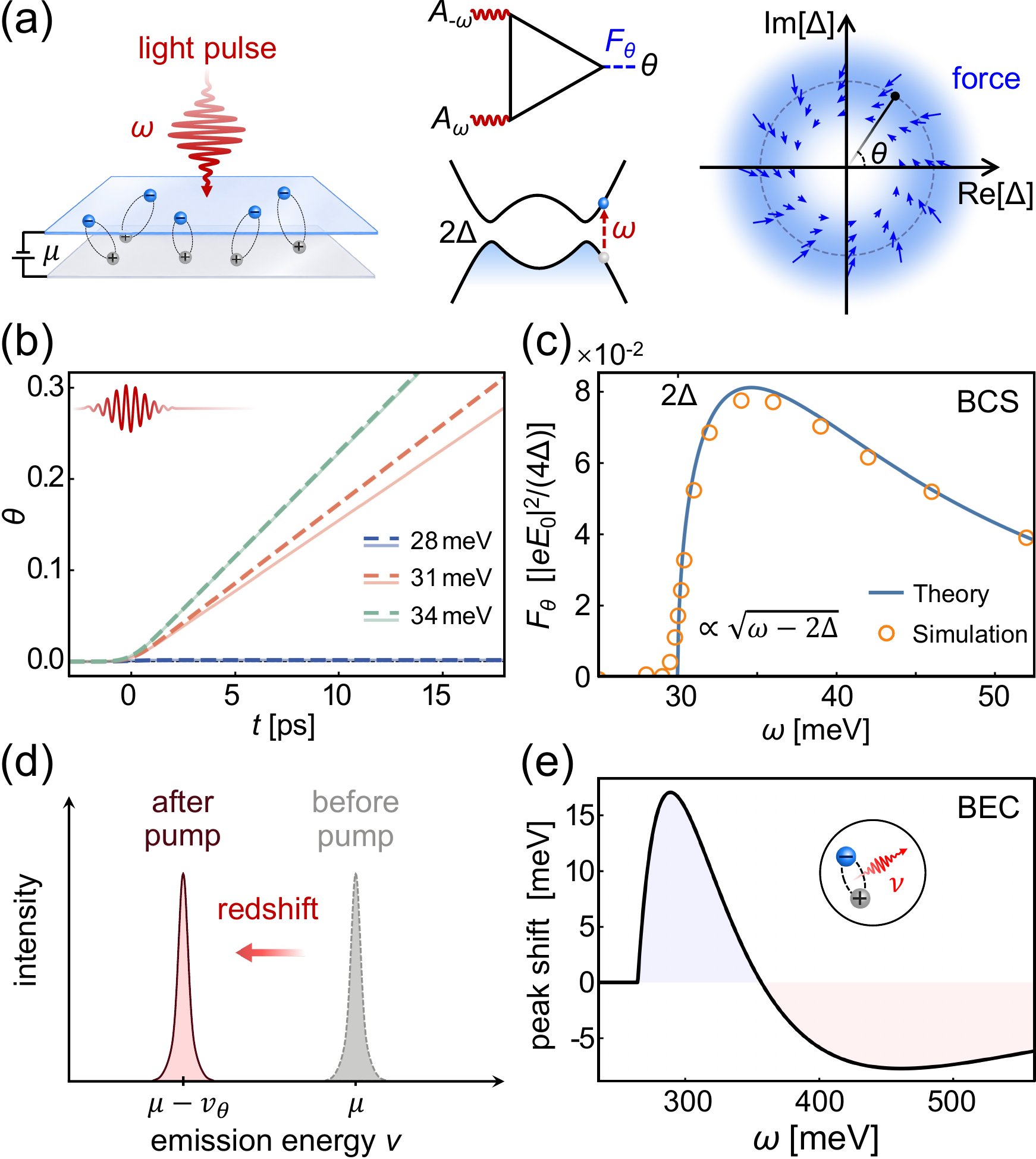}
  \caption{
  (a) Left: schematic of a light pulse driving an electron-hole bilayer realization of EI.
  Middle: the triangular diagram for the light-induced force $F_\theta$ on the order parameter phase (top) and the resonant interband excitation (bottom) responsible for it.
   Right:  light-induced force (arrows) on the order parameter shown in the complex order-parameter plane, where the color map shows the free-energy landscape without the drive. 
  (b) Simulated evolution of the phase in a BCS-type EI after a light pulse (inset) centered at time zero for three different central frequencies, with the same pulse duration $\tau=1~\mathrm{ps}$ and peak field $E_0=300~\mathrm{V/cm}$. 
  The solid/dashed curves show the results with fixed/dynamical amplitudes.
  (c) $F_\theta$ in a BCS-type EI as a function of the  photon energy. 
  The blue curve is the analytical result from Eq.~(\ref{ei_force2}) while the orange circles are from direct simulations. 
   The parameters for (b,c) are $E_{\rm G}=-380~\mathrm{meV}$, $m=0.2m_e$, and $\Delta=15~\mathrm{meV}$. 
  (d) An optical pump pulse  shifts the luminescence peak of a bias-sustained EI from $\mu$ to $\mu-v_\theta$. 
  (e) Shift of the luminescence peak of a BEC-type EI  plotted as a function of the pump frequency with fixed pulse duration $\tau=1~\mathrm{ps}$ and peak field $E_0=2\times10^5~\mathrm{V/cm}$.
  The electronic effective mass is $m=0.44m_e$ and the dielectric constant is $\epsilon=7$, corresponding to  the binding energy $E_{\rm b}=244~\mathrm{meV}$ and exciton Bohr radius $a_0=1.7~\mathrm{nm}$, consistent with \ce{MoSe2}/\ce{WSe2} double layers~\cite{Wang2019HighTemperatureExcitonCondensation, Ma2021StronglyCorrelatedEI,Nguyen2025PerfectCoulombDrag,Qi2025PerfectCoulombDrag}.
  To overcome the large original gap $G=1.74~\mathrm{eV}$, the bias voltage is $\mu=1.52~\mathrm{eV}$ which leads to  the exciton density $\rho_0=5.5\times10^{11}~\mathrm{cm^{-2}}$~\cite{Sun2024exciton}.
}    
  \label{fig:ei_phase}
\end{figure}

\emph{Excitonic Insulators---}We now introduce the main result: light-induced NCSF on the phase of the order parameter in excitonic insulators (EI), see Fig.~\ref{fig:ei_phase}(a).
An EI is a correlated state formed by condensed excitons that spontaneously breaks either the $U(1)$ symmetry of exciton number conservation~\cite{keldysh1965possible,
jerome1967excitonic,
kaneko2025new}, 
or certain discrete symmetries~\cite{
Portengen.1996,
zenker2014fate,
mazza2020nature,
golez2020nonlinear,
Sun.2021_Josephson,
kaneko2021bulk,
kaneko2025new}.
The simplest EI  is described by 
the Hamiltonian density~\cite{Sun_BaSh_2020, Dai_EI_2024} 
\begin{equation}
H=\Psi^\dagger
\mqty(\xi_{\vb{p}+e\vb{A}} & \Delta \\ \Delta^* & -\xi_{\vb{p}+e\vb{A}})\Psi
+\frac{1}{g}\abs{\Delta}^2
\,.
\label{ei_lagrangian}
\end{equation}
Here $\vb{p}=-i\nabla$ and  
$\vb{A}(t)=A_0 \hat{x} \sin\omega t $ is the vector potential of the coherent light encoding the electric field $\vb{E}(t)=-E_0 \hat{x} \cos\omega t $ where $E_0=\omega A_0$.
$\Psi=(\psi_c,\psi_v)^T$ are the annihilation operators for the bare 
conduction and valence bands with dispersion $\pm\xi_{\vb{k}}$, and  
 $\xi_{\vb{k}}=k^2/(2m)+E_{\rm G}/2$ with $E_{\rm G}$ being the band gap. 
We have set the Planck constant $\hbar$ and speed of light $c$ to unity for notational simplicity.
\equa{ei_lagrangian} is obtained after Hubbard-Stratonovich decoupling of the electron-hole  interaction $g$ by the complex field $\Delta$~\cite{Sun_BaSh_2020, Sun.2021_Josephson}.
In the EI phase,  $\Delta$ develops an expectation value, the excitonic order parameter  that spontaneously breaks the $U(1)$ symmetry of phase rotation.
It opens a quasiparticle gap in the BCS regime ($E_{\rm G} <0$, see Fig.~\ref{fig:ei_phase}(a)) and enhances the gap in the BEC regime ($E_{\rm G} >0$).
In the following, we take the ground state $\Delta$ to be a real positive number without loss of generality.


By integrating out the fermions, one obtains the Keldysh action~\cite{keldysh-ori,keldysh-mdn} that encodes  the effective dynamics of the order parameter $(\Delta+\delta) e^{i\theta}$ coupled to the optical field, where $\theta$ and $\delta$ are the phase and amplitude fluctuations. 
For time scales much longer than the inverse gap, the dynamics  is
\begin{subequations}\label{eq:bec-saddle-point}	
\begin{align}
\frac{\nu}{4} \partial_t^2\theta - C_0\partial_t\delta &=F_{\theta}
\,,
\\
 \nu \qty(\frac{1}{4\Delta^2}\partial_t^2+1)\delta +C_0\partial_t\theta &=F_{\delta}
\,,
\end{align}
\end{subequations}
see SI Sec.~III~\cite{supp} for the derivation.
Here, $\nu$ is the phase mode inertia and $C_0$ is the coefficient of linear phase-amplitude coupling.
In the BCS weak-coupling limit, $C_0$ vanishes and $\nu$ reduces to the	 bare electronic density of states.
The term $F_\theta \propto A_0^2$ is the NCSF on the phase that is induced by  resonant interband excitations when the light frequency $\omega$ exceeds the absorption threshold.
$F_{\delta}$ is the conservative force on the amplitude direction~\cite{sun2024floquet}.
A typical force field $(F_\theta, F_{\delta})$ is plotted  in \fig{fig:ei_phase}(a) where its nonconservative nature is apparent.
After eliminating the amplitude fluctuation, the low-energy dynamics of the phase is
\begin{equation}\label{eqn:EI_phase_EOM}
  \frac{1}{4}\tilde{\nu}\partial_t^2\theta 
  =  F_{\theta} + \frac{C_0}{\nu}\partial_t F_{\delta}
\,
\end{equation}
where $\tilde{\nu}=\nu(1+4C_0^2/\nu^2)$.

From the triangular diagram in \fig{fig:ei_phase}(a), the NCSF on the phase is derived as
\begin{equation}\label{ei_force}
F_\theta=\frac{\pi}{2}e^2\abs{A_0}^2\sum_{\vb{k}}\frac{v_{kx}^2\xi_{\vb{k}}\Delta^2}{E_{\vb{k}}^3}\delta(\omega-2E_{\vb{k}})
\,
\end{equation}
where  $E_{\vb k}=\sqrt{\xi_{\vb k}^2+\Delta^2}$ and $v_{kx}=\partial \xi_{\vb k}/\partial k_x$, see SI Sec.~IIIB~\cite{supp} for details.
This result may be understood in terms of driven precession  of Anderson pseudospins.
The pseudospin ${\vb s}$ at momentum ${\vb k}$ precesses under the pseudomagneic field ${\vb B}=-(\Delta,0,\xi_{\vb k + e \vb A(t)})$.
At second order in $A$, a nonzero static value of ${s}_{y}$ at this momentum emerges, and contributes a force on the phase that is being summed in \equa{ei_force}, see Appendix B.
The delta function indicates that this force relies on resonant interband excitations.  
Because of the $\xi_{\vb{k}}$ term, the contributions from the excitations on the `particle side' ($\xi_{\vb{k}}>0$) and the `hole side' ($\xi_{\vb{k}}<0$) tend to cancel each other, so that the force vanishes in the ``particle-hole'' symmetric limit.
In two dimensions (2D) with quadratic dispersion of bare carriers, the bare density of states  is a constant, while the energy-dependent velocity $v_{kx}$ leads to an imbalance between `particle side'  and `hole side' contributions, so that a nonzero $F_\theta$ exists.
Because of the $U(1)$ symmetry of varying the phase, the force must be constant along this compact coordinate, and is therefore nonconservative.

In the weak-coupling BCS limit ($E_{\rm G}<0$, $\Delta\ll |E_{\rm G}|$) and the dilute BEC limit ($E_{\rm G}>0$, $\Delta\ll E_{\rm G}$), 
closed-form expressions for the force are found from \equa{ei_force} in  2D as
\begin{equation}
\begin{aligned}
F_\theta=
\frac{e^2\abs{E_0}^2}{4\Delta}
\left\{
\begin{array}{ll}
\displaystyle\frac{2\Delta^3 \sqrt{\omega^2-4\Delta^2}}{\omega^4}\Theta(\omega-2\Delta), & \text{BCS},
\\
\displaystyle\frac{\Delta^3(\omega-\tilde{E}_{\rm G})}{\omega^4}\Theta\qty(\omega-\tilde{E}_{\rm G}), & \text{BEC}.
\end{array}
\right.
\end{aligned}\label{ei_force2}
\end{equation}
Here $\tilde{E}_{\rm G}=\sqrt{4\Delta^2+E_{\rm G}^2}$ is the optical gap in the BEC regime.   
Obviously, the force is nonzero only when the light frequency is above the optical gap, as shown in Fig.~\ref{fig:ei_phase}(c). 
Fig.~\ref{fig:ei_phase}(b) shows direct numerical simulations of the phase evolution of a 2D EI after being excited by a light pulse, which confirm 
that above-gap pulses produce a nonzero phase winding velocity. 
This distinguishes the NCSF from conventional Floquet band renormalization, which disappears together with the drive and does not leave persistent phase winding.

For an EI with strictly conserved exciton number, the absolute global phase is not directly observable. 
However, in most excitonic insulators such as those in natural solids, the
exciton number is not conserved,
and the phase of the order parameter is often associated with physical observables such as electrical polarization~\cite{Portengen.1996, Sun.2021_Josephson} or lattice distortion~\cite{mazza2020nature}.
Therefore, the NCSF could be measured by monitoring these observables after a pump
pulse~\cite{Ning.2020_TNS, Liu2021PhotoinducedTa2NiSe5, Haque2024TerahertzParametric}.
Here we propose the most direct way to observe this NCSF, which relies on the luminescence of  excitonic condensates such as those  in biased electron-hole bilayers~\cite{Sun2024exciton, Wang2019HighTemperatureExcitonCondensation, Ma2021StronglyCorrelatedEI,Nguyen2025PerfectCoulombDrag,Qi2025PerfectCoulombDrag} shown schematically in Fig.~\ref{fig:ei_phase}(a).
Before the pump, the phase of the condensate  is already evolving as $\Delta=\abs{\Delta}e^{-i\mu t}$ with $\mu$ set by the interlayer bias voltage. 
Weak exciton recombination emits coherent photons  at the energy $\mu$. 
A pump pulse of duration $\tau$ induces the NCSF, which leads to an additional winding speed 
$v_\theta \approx \tau \partial_t^2\theta $ 
of the phase following \equa{eqn:EI_phase_EOM}.
In turn, it shifts the emitted photon energy by
$-v_\theta$, which could be measured by time-resolved detection of the luminescence, see \fig{fig:ei_phase}(d).

A likely near-term realization of EI is the BEC-type formed by excitons bound by Coulomb attraction~\cite{Sun2024exciton,Wang2019HighTemperatureExcitonCondensation, Ma2021StronglyCorrelatedEI,Nguyen2025PerfectCoulombDrag,Qi2025PerfectCoulombDrag,qi2026nature}.
There, the momentum dependence of the order parameter $\Delta_{\vec{k}}$ has to be accounted for when computing $F_{\theta}$ and  $\partial_t F_{\delta}$.
After a light pulse of duration $\tau$ and peak field $E_0$, the change of phase winding velocity is found as
\begin{align}
v_\theta
&\approx
16 \pi^{\frac{5}{2}} 
\tau \abs{eE_0}^2  
\frac{\rho_0 E_{\rm b}^2 (\omega-\tilde{E}_{\rm G})}{m^2\omega^5}
\left(
1
-
\frac{2 E_{\rm b}^2}{\omega^2}
\right)
\Theta\qty(\omega-\tilde{E}_{\rm G}) 
\,,
\label{ei_experiment}
\end{align}
see SI Sec.~IIID~\cite{supp}.
Here $\rho_0$ is the exciton condensate density and $E_{\rm b}$ is the binding energy.
Note that the optical gap $\tilde{E}_{\rm G}$ for exciton-breaking excitations is set by the bare gap $E_{\rm G}=G-\mu$  that accounts for the voltage bias, not the original large gap $G$.
This phase winding manifests directly as the shift of the photon emission energy, which can either be a blue or red shift depending on the pump frequency relative to $\sqrt{2}E_{\rm b}$, see Figs.~\ref{fig:ei_phase}(d)(e).
The blue shift is caused by the $\partial_t F_{\delta}$ term in \equa{eqn:EI_phase_EOM}: optical excitation provides a force that tends  to reduce the exciton density, which acts in the same way as raising the exciton energy.
The red shift comes from $F_{\theta}$, which reflects that after pair-breaking excitation reduces the number of excitons, the average energy of an exciton is lowered because of reduced repulsion.
We emphasize that despite the physical understanding, the NCSF framework is needed to determine the shift quantitatively, especially in the BCS regime where the single exciton picture is no longer clear.

For representative \ce{MoSe2}/\ce{WSe2} double layers~\cite{Wang2019HighTemperatureExcitonCondensation, Ma2021StronglyCorrelatedEI,Nguyen2025PerfectCoulombDrag,Qi2025PerfectCoulombDrag}, the shift is of order $5 \unit{meV}$ after a $\tau=1\unit{ps}$ pulse with $E_0=2\times10^5~\mathrm{V/cm}$, see \fig{fig:ei_phase}(e). 
Contrary to the common mechanisms such as the optical Stark effect (ponderomotive shift) that occurs only during the pump pulse, the NCSF-induced shift has a threshold behavior at $\omega=\tilde{E}_{\rm G}$, and persists over the exciton number relaxation time (often exceeding nanoseconds~\cite{Rivera2015LongLived,Jauregui2019Electrical}) after the pump pulse.
The frequency-dependent sign reversal in Figs.~\ref{fig:ei_phase}(e) also distinguishes this effect from simple heating.

\begin{figure}
	\includegraphics[width=\linewidth]{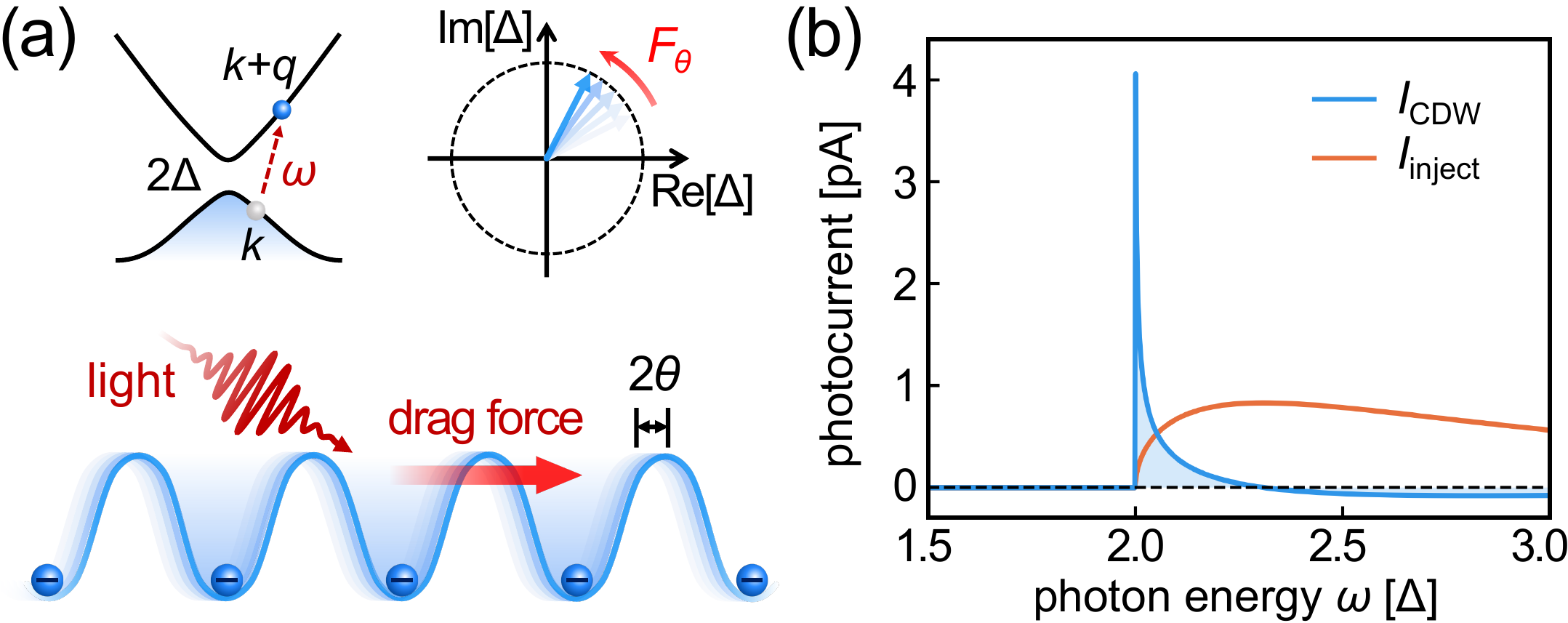}
	\caption{
		(a) Bottom panel illustrates the light-induced drag of a one-dimensional CDW with electrons (spheres) sitting in the periodic mean-field potential (cyan curve). 
		The phase $\theta$ of  the CDW order parameter sets the spatial position of the condensate. 
		The top left depicts the resonant  interband excitations responsible for the drag force $F_\theta$. The top right shows this force on the complex plane of the order parameter.
		(b) 
		The blue curve is the  NCSF-induced photo-current from CDW sliding as a function of light frequency (in units of the CDW gap $\Delta$) and the orange curve is the quasiparticle injection current.
		The obliquely incident p-polarized light has a fixed  electric field $E_0=10^4~\mathrm{V/cm}$ and incident angle $45^\circ$.
		The other parameters are $\Delta=75~\mathrm{meV}$, $k_{\rm F}=3.1~\mathrm{nm^{-1}}$, $m=m_e$, $\nu^*/\nu_0=100$, $\gamma=1~\mathrm{THz}$, $\gamma_{\rm tr}=25~\mathrm{THz}$, chosen to be representative of  \ce{K_{0.3}MoO3}~\cite{travaglini1984, demsar1999,guster2019, kang2021nc}.
		The plotted current is for a  single chain.
        With one chain per $\unit{nm^2}$ of cross-sectional area, a $10 \unit{\mu m}$-wide and $10 \unit{nm}$-thick  sample would give a photocurrent $\sim 0.1 \unit{\mu A}$.
	}\label{fig:cdw} 
\end{figure}

\emph{\label{sec:cdw}Charge Density Waves---}We now move on to the second nontrivial example, CDW systems arising from Fermi surface nesting shown in Fig.~\ref{fig:cdw}(a). 
We consider the minimal one-dimensional spinless Peierls model along the $x$ direction with the  Lagrangian
\begin{equation}\label{eqn:H_CDW}
L=\bar{\Psi}
  \mqty(
  -i\partial_t + \xi_{p} & \Delta \\
  \Delta^\ast  &  -i\partial_t + \xi_{-p}
  )
  \Psi
  +\frac{|\Delta|^2}{2g}
  - \frac{|\dot{\Delta}|^2}{2g \omega_Q^2}
  ,
\end{equation}
where the fermion fields $\Psi=(\psi_{\rm R}(x),\psi_{\rm L}(x))^T$ are  for
the right (R) and left (L) moving electrons around the two Fermi points  shifted by the Fermi momentum $\mp k_{\rm F}$ to the Gamma point,
exhibiting the band dispersions 
$\xi_k=v_{\rm F} k + k^2/(2m)$ and $\xi_{-k}$.
The EM vector potential $A(x,t)$ couples to electrons via $p+eA$.
A uniform ground state CDW order parameter $\Delta=|\Delta|e^{i2\theta}$ is defined from the lattice distortion $u(x)=\Delta e^{i 2k_{\rm F} x} + \rm c.c.$ that modulates the ionic and electronic density at the wave vector $Q=\pm 2k_{\rm F}$. 
The $1/(2g)$ in the second term denotes the stiffness of the relevant lattice phonon mode against distortion and the last term is the kinetic energy of this mode with $\omega_Q$ being its bare eigenfrequency.

Despite the similarity between the CDW in \equa{eqn:H_CDW} and the EI in \equa{ei_lagrangian}, the force in \equa{ei_force} \emph{does not} exist for CDW under uniform optical field because of  subtle differences.
In the CDW, a change of the phase $\theta$ translates the density wave in space by $\delta x=\delta\theta/k_{\rm F}$, see \fig{fig:cdw}(a). 
Since uniform light respects inversion symmetry after time averaging, it cannot exert a static force on the phase.
Indeed, the triangular diagram in \fig{fig:ei_phase}(a) for the CDW system yields \equa{ei_force} with $\xi_k$ replaced by $v_{\rm F} k$, where one can see that 
inversion symmetry leads to exact cancellation between excitations at $k$ and $-k$.
Nevertheless, if a propagating electromagnetic field $A(x,t)=A_0\sin(\omega t-qx) \hat{x}$ with a nonzero wave vector $q$ is applied, it selects a direction and allows a rectified drag force on the  phase. 
After integrating out the electrons and the fast order parameter oscillations, the slow ($\dot{\theta} \ll 2\Delta, \omega$) dynamics for the phase is obtained as
\begin{equation}
  \nu^\ast (\partial_t^2+\gamma\partial_t)\theta = F_{\theta} ,
\label{cdw_phase_eom}
\end{equation}
see SI Sec.~IV~\cite{supp}.
Here
$
\nu^\ast=\nu_0+ 4|\Delta|^2/(g \omega_Q^2)
$ 
is the inertia of the phase contributed by  bare electronic density of states $\nu_0$ and the lattice mode.
We have also added a damping rate $\gamma$ arising from lattice friction. 
Note that because of inversion symmetry, the phase mode is decoupled from the amplitude mode at linear order, so that the amplitude mode effect in \equa{eqn:EI_phase_EOM} does not appear here.

From the triangular diagram in \fig{fig:ei_phase}(a), with the optical vertices being the propagating field plus the induced linear-response phase oscillation, the NCSF is derived to leading order in $q$ as
\begin{align}\label{eqn:CDW_force}
F_\theta
&= \frac{q}{v_{\rm F} m  \omega}P_{\rm qp}(\omega)-\frac{1}{v_{\rm F}}K(\omega)\abs{E_0}^2
\,,
\end{align}
see SI Sec.~IVB~\cite{supp}. Here $v_{\rm F}$ is the Fermi velocity and $E_0=\omega A_0$ is the driving electric field.
The first term has a simple interpretation: $P_{\rm qp}$ is the absorption power of light due to interband excitation of quasiparticles.
The factor $q/\omega$ converts it into the total in-chain momentum of absorbed photons, which provides a force to push the CDW to slide. 
This connection becomes particularly transparent when the lattice friction $\gamma$ vanishes, for which quasiparticle excitation is the only source of dissipation and $P_{\rm qp}$ is directly set by the real part of the  optical conductivity $\Re[\sigma(\omega)]$, reducing Eq.~\eqref{eqn:CDW_force} to 
$F_\theta=v_{\rm F}^{-1}[q\Re[\sigma]/(2m\omega)-K]|E_0|^2$.
Note that part of the absorbed momentum is  transferred to quasiparticle motion, which is why a second term, $K(\omega) \propto q$, is subtracted.
The excited quasiparticles contribute a nonlinear ``injection current''
$\partial_t I_{\rm inject}\propto K(\omega)E_0^2$~\cite{justin2021}. 
With a transport scattering rate $\gamma_{\rm {tr}}$, the steady state injection current is $I_{\rm inject}= eK(\omega)E_0^2/\gamma_{\rm {tr}}$.
In the large-ion-mass regime ($\nu^* \gg \nu_0$), the force is simply
\begin{align}\label{eqn:CDW_force2}
F_\theta
\approx \frac{q e^2 \abs{E_0}^2}{m\omega^2}
\frac{\Delta^2 }{\omega \sqrt{\omega^2 - 4\Delta^2}}
\left(
\frac{16\Delta^2}{\omega^2}
-3
\right)
\Theta(\omega-2\Delta)
\,
\end{align}
which is strongly enhanced right above the optical gap.

For a long pulse or continuous-wave drive, a static $F_\theta$ pushes the phase to increase with the speed $\dot{\theta} = F_\theta/(\nu^* \gamma)$ in the steady state, creating a sliding CDW as a limit-cycle state shown in Fig.~\ref{fig:cdw}(a).
Since the increase of phase translates the ionic density wave in space by $\delta x=\delta\theta/k_{\rm F}$, the electrons are translated together with it.
It leads to a topological DC electric current $I_{\rm CDW} = e \dot{\theta} /\pi = e F_\theta /(\pi \nu^\ast \gamma)$ 
contributed by the condensate
via Thouless pumping~\cite{Thouless1983Quantization}.
The effects that modify $I_{\rm CDW}$ such as heating and non-adiabatic sliding occur only at higher orders in the optical field.
The NCSF thus predicts a nontrivial contribution to the photo-drag DC current in a CDW system.
Note that in contrast to conventional photon drag, where momentum transfer acts on mobile carriers, here the drag force acts on the collective order parameter of a broken-symmetry state.
The NCSF framework provides a microscopic prediction for this collective photocurrent.

The simplest experiment to observe this effect is to measure the photon-drag current in quasi-one-dimensional CDW systems such as \ce{ZrTe3}~\cite{zrte3-se-prl},
\ce{NbSe3}, \ce{TaS3}
and the incommensurate phase of \ce{K_{0.3}MoO3}~\cite{travaglini1984, demsar1999,guster2019, kang2021nc, pouget2021}.
As shown in Fig.~\ref{fig:cdw}(b), $I_{\rm CDW}$ rises sharply as the photon energy $\omega$ goes above the interband absorption edge $2\Delta$, reflecting that the NCSF relies on interband absorption.
A surprising prediction of  \equa{eqn:CDW_force2} is that $I_{\rm CDW}$ reverses its sign beyond  $\omega=4\Delta/\sqrt{3}$ .
The sharp peak of NCSF-induced $I_{\rm CDW}$ is the strongest feature to  distinguish it from the quasiparticle injection current $I_{\rm inject}$, also shown in Fig.~\ref{fig:cdw}(b).
Note that pinning from disorder generally exists~\cite{rice1976,gruner1988}, so that the incident light needs to exceed a certain threshold to induce the CDW sliding, offering another way to distinguish the NCSF effect from $I_{\rm inject}$.
Finally, the force would be greatly enhanced if the incident `light' is  a propagating polariton wave~\cite{Basov2016PolaritonsVDW, Low2017Polaritons2DMaterials} from, e.g., an adjacent hBN flake, which has much larger momentum than free-space photons.


To conclude, we note that symmetry and topology of the manifold of slow degrees of freedom impose general constraints on the effective static force, offering guidance for identifying NCSFs, see Appendix~C.
In addition to the effective static force, the periodic drive also generates effective dissipation (friction under slow motion) and fluctuations (noise) for the slow degrees of freedom, which warrant future investigation.
With these ingredients, one obtains a low-energy field theory for optically driven systems in the form of a Keldysh path integral, see Appendix~A.
It has the capability of predicting nonequilibrium steady states such as limit cycles and robust time crystals~\cite{Yao.2023,Diehl.2025_time_crystal} beyond the mean-field level.


\begin{acknowledgments}
This work is supported by the National Natural Science Foundation of China (Grants No. 12421004 and No. 12374291), Beijing Natural Science Foundation (Z240005), and the National Key Research and Development Program of China (2022YFA1204700). 
We thank T. Xiao, M. M. Fogler, A. J. Millis and J. Zhang for helpful discussions.
\end{acknowledgments}

\emph{Data availability---}Numerical codes and data for plots in this paper are available online [to be inserted].

$^*$ These authors contributed equally to this work.

$^\dagger$ Corresponding author:  zysun@tsinghua.edu.cn


\bibliography{NCSF}

\onecolumngrid
\begin{center}
{\large\textbf{End Matter}}
\end{center}
\twocolumngrid

\appendix

\setcounter{equation}{0}
\renewcommand{\theequation}{A\arabic{equation}}
\renewcommand{\theHequation}{A.\arabic{equation}}

\emph{Appendix A: Keldysh formalism for the effective force---}The Keldysh path integral description of the driven (closed or open) system is
\begin{align}\label{GeneratingFunctional}
  Z&=\int D[X_{\text{cl/q}},\phi_{\text{cl/q}}]e^{-iS[X_{\text{cl/q}},\phi_{\text{cl/q}};f]}
 \notag\\
 & =\int D[\phi_{\text{cl/q}}]e^{-iS_{\rm eff}[\phi_{\text{cl/q}};f_0]}
\end{align}
defined on the closed time contour~\cite{keldysh-mdn,altland2010condensed,keldysh-ori,schwinger1961brownian}.
Here the action is 
$S=\int dt L$ and $L[X_{\text{cl/q}},\phi_{\text{cl/q}};f]$ is the Keldysh Lagrangian. 
The $X$ denotes the fast degrees of freedom whereas $\phi$ denotes the slow degrees of freedom.
The subscripts `cl' and `q' denote the `classical' and `quantum' components of the fields,  following the Keldysh notation.
For example, $X_{\rm cl}=(X_{+}+X_{-})/2$ and $X_{\rm q}=X_{+}-X_{-}$  are the `classical' and `quantum'  components of  $X$, while $X_{+}(t)$ and $X_{-}(t)$ are their values on the forward and backward branches of the time contour. 
The $f(t)=f_0 \cos \omega t$ is the oscillating external driving field.
Integrating out the fast fields $X$ results in
 the low-energy effective action 
$S_{\rm eff}=\int dt L_{\rm eff}$ where $L_{\rm eff}$ is
the effective Lagrangian for the slow field.
It can be expanded to leading order in $\phi_{\rm q}$ as
\begin{align}
L_{\rm eff}[\phi_{\text{cl/q}};f_0] 
= & L_s[\phi_{\text{cl/q}}] - \phi_{\rm q} F_{{\rm eff}}(\phi_{\rm cl};f_0) 
+\cdots
\label{eq:effective_force_action}
\end{align}
where $F_{{\rm eff}}$ is the (drive-induced) effective static force acting on $\phi$, 
as implied by the classical equation of motion $\delta S_{\rm eff}/ \delta \phi_{\rm q} =0$.
The higher-order terms in $O(\partial_t\phi_{\rm q},\phi_{\rm q}^2)$ encode dissipation (frictional forces) and fluctuations (noise).
Note that $\phi$ represents multiple degrees of freedom and $\phi_{\rm q} F_{{\rm eff}}$ should be viewed as a dot product between $\phi_{\rm q}$ and the multi-component force $F_{{\rm eff}}$.

\emph{Relation of NCSF to dissipation---}A generic form of the original Lagrangian may be written as
\begin{equation}\label{eqn:generic_L}
	L= L_{\rm f}[X_{\text{cl/q}},\phi_{\text{cl/q}}]+L_s[\phi_{\text{cl/q}}]+L_{\rm c}[X_{\text{cl/q}},\phi_{\text{cl/q}},t]
\,
\end{equation}
where $L_{\rm c}[X_{\text{cl/q}},\phi_{\text{cl/q}},t]=L_{c0}[X_{+},\phi_{+},t]-L_{c0}[X_{-},\phi_{-},t]$ is the time-periodic driving term.
Ref.~\cite{sun2024floquet} proved that the effective static force $F_{{\rm eff}}(\phi, f_0)$ in \equa{eq:effective_force_action} must be the gradient of an effective potential, the ponderomotive potential $V_{\rm P}(\phi, f_0)$, in two cases.
Case~1 is the dissipationless case, meaning that there is no resonant excitation of the system by the oscillating drive and no level broadening from a bath. 
Case~2 is for a special form of the Lagrangian:
$L_{\rm f}[X_{\text{cl/q}}]$ does not depend on the slow field $\phi_{\text{cl/q}}$, and the driving term is separable  as $L_{c0} = P_1(X) P_2(\phi) \cos(\omega t)$.

The classical limit of case~1 is particularly transparent. For a closed system,
the ponderomotive force on $\phi$ implied by \equa{eqn:generic_L} is 
\begin{align}
F_{\rm eff}=&-\langle \partial_\phi L_{\rm {fc}} \rangle
=- d_\phi \langle L_{\rm {fc}} \rangle 
\nline
&
+
\langle 
\frac{\delta X}{\delta \phi}  
\left(
\partial_X L_{\rm {fc}} 
-
d_t \partial_{\dot{X}} L_{\rm {fc}} 
\right)
\rangle
+
\langle 
d_t\left(\frac{\delta X}{\delta \phi}   \partial_{\dot{X}} L_{\rm {fc}} 
\right)
\rangle
\nline
=&
- d_\phi \langle L_{\rm {fc}} \rangle +
\langle 
d_t\left(\frac{\delta X}{\delta \phi}   \partial_{\dot{X}} L_{\rm {fc}} 
\right)
\rangle
\end{align}
where $L_{\rm {fc}}=L_{\rm f}+L_{\rm c}$ and the average is taken over a time interval  $T \gg 1/\omega$.
Note that as $\phi$ varies adiabatically, the classical solution for the fast field $X[t,\phi]$ also varies.
The last equality is obtained by noting that $\partial_X L_{\rm {fc}} 
-d_t \partial_{\dot{X}} L_{\rm {fc}} =0$ from the equation of motion.
The last term is a boundary term since it is a total derivative.
If there is no dissipation (no optical absorption on average), the dynamics of $X$ must be bounded, and the boundary term vanishes as long as $T$ is large.
Therefore, one concludes that the ponderomotive force is conservative, and the corresponding potential is simply the time-averaged Lagrangian $\langle L_{\rm {fc}} \rangle $~\cite{sun2024floquet}.


Therefore, for $F_{{\rm eff}}$ to have a nonconservative component,  dissipation is a necessary condition.
The connection between the nonconservative effective force and dissipation is
explicit in a typical example (a slight generalization of `case~2') of \equa{eqn:generic_L}:
\begin{align}
& L_{\rm f}[X_{\text{cl/q}},\phi_{\text{cl/q}}] =L_{\rm f}[X_{\text{cl/q}}]
,\notag \\
& 
L_{\rm c0}[X,\phi,t]=E(\phi,t) P[X],
\label{eq:classical_lagrangian}
\end{align}
where the slow field  affects the fast one only through the driving field $E(\phi,t)$ that couples to the generalized polarization $P$.
Expanding the coupling to linear order in $ \phi_{\rm q} $ gives
\begin{equation}
L_{\rm c} = P_{\rm q}  E(\phi_{\rm cl},t)
+ P_{\rm cl}  \phi_{\rm q} 
\partial_{\phi_{\rm cl}}E(\phi_{\rm cl},t)
,
\label{eq:keldysh_coupling}
\end{equation}
where $P_{\rm cl}=(P[X_+]+P[X_-])/2$ and $P_{\rm q}=P[X_+]-P[X_-]$. 
A generic drive field could be written as 
$E(\phi,t)=E_0(\phi)\cos[\vartheta(\phi)-\omega t]$ where the slow field $\phi$ affects not only its amplitude but also its phase $\vartheta(\phi)$.
The latter is the key to induce a NCSF.
This model covers the photo-drag class, see SI Sec.~I~\cite{supp}.
Integrating out $X$ in \equa{eqn:generic_L} with the terms from \equa{eq:classical_lagrangian} yields the effective force on $\phi$.

From \equa{eq:keldysh_coupling}, a contribution to the force is 
$\hat{F}_{\rm NCSF}=(\partial_\phi\vartheta) P \partial_t E/\omega$. Its path integral average over fast field therefore gives $F_{\rm NCSF}=\langle P \partial_t E\rangle \partial_\phi\vartheta /\omega=-\langle E \partial_t P \rangle \partial_\phi\vartheta /\omega=(\partial_\phi \vartheta) \mathcal{P}_{\rm in}/\omega$ where $\mathcal{P}_{\rm in}=-\langle E \partial_t P \rangle$ is exactly the injected power of the drive to the system.
If $\phi$ is a single variable for a closed  manifold: $\phi \equiv \phi+2\pi$, the map from $\phi$ to the phase $\vartheta(\phi)$ can have a nonzero winding number. 
In this case, it is obvious that $F_{\rm NCSF}$ is nonconservative.
For example, a traveling drive  in $\phi$ space  has $\vartheta=\kappa \phi$ where $\kappa \in \mathbb{Z}$ is the winding number. 
The total force is found as
\begin{align}
F_{\rm eff} 
&= -\partial_\phi V_{\rm P} + F_{\rm NCSF},
\label{eq:Feff}
\\
V_{\rm P}(\phi) &= -\sum_{n=1}^\infty A_n \Re\qty[\chi^{(2n-1)}_{\rm R}]E_0(\phi)^{2n}
\,,
\label{eq:V_P}
\\
F_{\rm NCSF} &= \sum_{n=1}^\infty B_n \Im\qty[\chi^{(2n-1)}_{\rm R}]E_0(\phi)^{2n} \partial_\phi\vartheta \,,
\label{eq:NCSF_force}
\,
\end{align}
where $\chi^{(2n-1)}$ are retarded correlation functions of $P[X]$ in equilibrium: 
\begin{align}
\chi^{(1)}_{\rm R} &=-i\expval{P_{\rm cl}(0) P_{\rm q}(t)}|_\omega,
\nline
\chi^{(3)}_{\rm R} &=(-i)^3\expval{P_{\rm cl}(0) P_{\rm q}(t_1) P_{\rm q}(t_2)P_{\rm q}(t_3)}|_{\omega; \omega, -\omega, -\omega},
\nline
&\cdots
\label{eq:chi_R}
\,
\end{align}
where $A_n=\frac{1}{2^{2n}(n!)^2},\, B_n=2nA_n$ are combinatorial factors.
These correlation functions ($\chi^{(2n-1)}_{\rm R}$) are just linear ($n=1$) and nonlinear ($n>1$) response functions of $P$ in equilibrium~\cite{keldysh-mdn,altland2010condensed}.
Note that the $\chi^{(2n-1)}_{\rm R}$ is the average of all permutations with the frequency arguments summing to zero while that in  the second line of \equa{eq:chi_R} just shows one permutation.
The first term in \equa{eq:Feff} is conservative. It is the gradient of  the ponderomotive potential \equa{eq:V_P} that is related to the real parts of $\chi_R^{(2n-1)}$.
\equa{eq:NCSF_force} is the nonconservative force  originating from the dissipative parts (imaginary parts) of the response functions.
If $\kappa=0$, the drive reduces to `case~2' discussed above and only the $V_{\rm P}$ term survives, consistent with Eq.~(4) of Ref.~\cite{sun2024floquet}.

\setcounter{equation}{0}
\renewcommand{\theequation}{B\arabic{equation}}
\renewcommand{\theHequation}{B.\arabic{equation}}
\emph{Appendix B: Light-induced force on the phase of excitonic insulators---}
The derivation of the force  in terms of Green functions is contained in SI Sec.~III~\cite{supp}.
Here we show a simpler derivation in terms of the precession of `Anderson pseudospins'. 
The mean-field Hamiltonian in \equa{ei_lagrangian} may be written as $H=-\sum_{\vb k} \vb{B}_{\vb{k}} \cdot \Psi_{\vb k}^\dagger \sigma \Psi_{\vb k}$ 
where $\Psi_{\vb k}=(\psi_{c,\vb k},\psi_{v,\vb k})^T$ is the two-component annihilation operator at momentum $\vb k$, 
$\vb{B}_{\vb{k}}=-(\Delta,0,\xi_{\vb{k}})$ is the pseudomagnetic field at $\vb k$,
and $\sigma \equiv (\sigma_x, \sigma_y, \sigma_z) =(\sigma_1, \sigma_2, \sigma_3)$.
One defines the pseudo-spin $s_{\vb k}=\langle \Psi_{\vb k}^\dagger \sigma \Psi_{\vb k} \rangle/2$.
The EI mean-field Hamiltonian with a real ground state order parameter $\Delta$ aligns the pseudospins parallel to the pseudomagnetic field in the $x$-$z$ plane.
A phase fluctuation corresponds to a tilt toward the $s_y$ direction. 



The optical field generates time-dependent perturbations through the paramagnetic coupling 
$\vb{B}_{\vb{k}}^{(1)}=-(0,0,\vb{v}_{\vb{k}}\cdot\vb{A}(t))$ and diamagnetic coupling $\vb{B}_{\vb{k}}^{(2)}=-(0,0,A^2(t)/(2m))$. The pseudospin dynamics is governed by the Landau-Lifshitz-Gilbert equation
\begin{equation}
\partial_t{\vb{s}}_{\vb{k}} = 2\vb{s}_{\vb{k}} \times (\vb{B}_{\vb{k}}^{\rm tot} - 
\gamma  \partial_t{\vb{s}}_{\vb{k}} )  
\end{equation}
with $\vb{B}_{\vb{k}}^{\rm tot} = \vb{B}_{\vb{k}} + \vb{B}_{\vb{k}}^{(1)} + \vb{B}_{\vb{k}}^{(2)}$ and $\gamma$ being the Gilbert damping. 
We expand the solution order by order in the optical field $A$, $\vb{s}_{\vb{k}}(t)=\vb{s}_{\vb{k}}^{(0)}+\vb{s}_{\vb{k}}^{(1)}(t)+\vb{s}_{\vb{k}}^{(2)}(t)$.
The zeroth-order solution gives the equilibrium pseudospin $\vb{s}_{\vb{k}}^{(0)}=-\frac{1}{2E_{\vb{k}}}(\Delta,~0,~\xi_{\vb{k}})$ aligned parallel to 
$\vb{B}_{\vb{k}}$, corresponding to occupying the valence band in the EI state. 
The linear and second order responses satisfy
\begin{align}
-\partial_t{\vb{s}}_{\vb{k}}^{(1)} 
&= 2\vb{B}_{\vb{k}}^{(1)} \times \vb{s}_{\vb{k}}^{(0)}+ 2\vb{B}_{\vb{k}} \times \vb{s}_{\vb{k}}^{(1)}+ 2\gamma \vb{s}_{\vb{k}}^{(0)} \times \partial_t{\vb{s}}_{\vb{k}}^{(1)}
\,,
\\
-\partial_t{\vb{s}}_{\vb{k}}^{(2)} 
&= 2\vb{B}_{\vb{k}}^{(2)} \times \vb{s}_{\vb{k}}^{(0)}
+ 2\vb{B}_{\vb{k}}^{(1)} \times \vb{s}_{\vb{k}}^{(1)} 
+ 2\vb{B}_{\vb{k}} \times \vb{s}_{\vb{k}}^{(2)}
\nline
&\quad+ 2\gamma \vb{s}_{\vb{k}}^{(1)} \times \partial_t{\vb{s}}_{\vb{k}}^{(1)}
+ 2\gamma \vb{s}_{\vb{k}}^{(0)} \times \partial_t{\vb{s}}_{\vb{k}}^{(2)}
\label{eqn:LLG_2}
\,.
\end{align}
The linear-response solution is
\begin{align}
\vb{s}_{\vb{k}}^{(1)}(t)
&=
\frac{\Delta}{E_{\vb{k}}}
\left(
\frac{\xi_{\vb{k}}}{E_{\vb{k}}}
(2E_{\vb{k}}+\gamma\partial_t)
,\,
{
-\partial_t
},\,
{
-\frac{\Delta}{E_{\vb{k}}}
(2E_{\vb{k}}+\gamma\partial_t)
}
\right)
\nline
& \quad\times \frac{1}{\partial_t^2+(2E_{\vb{k}}+\gamma\partial_t)^2}
\vb{v}_{\vb{k}}\cdot\vb{A}(t)
\,
\end{align}
where the $\partial_t$ acts on $\vb{A}(t)$.
The effective forces are encoded in the static second-order component $\vb{s}_{\vb{k}}^{(2)}(\omega=0)$. 
Note that for a long (temporally broad) optical pulse, the static component of the $\vb{B}_{\vb{k}}^{(2)}$ term in \equa{eqn:LLG_2} can be viewed as a `static' field $\vb{B}_{\vb{k}}^{(2)}=-(0,0,A_0^2/(4m))$ that is slowly turned on.
This term induces an adiabatic precession of the pseudospins around the $z$ axis, with a phase angle approaching a constant in the steady driven state.
To find a steady-state solution, terms of this kind can be absorbed in the static pseudomagnetic field in a rotated frame so that it still lies within the $x-z$ plane.
Therefore, these terms do not contribute to the force on the phase direction, and will be neglected in the following.
Since the $\partial_t{\vb{s}}_{\vb{k}}^{(2)}$ terms are zero for the steady state, a nonzero $y$ component of the pseudospin is thus  found from \equa{eqn:LLG_2} to second order in the optical field  as 
\begin{align}\label{eqn:sy2}
\expval*{s_{{\vb{k}} y}^{(2)}}
&=\frac{\gamma}{\Delta}\expval{s_{{\vb{k}} x}^{(1)}\partial_t s_{{\vb{k}} y}^{(1)}
		-s_{{\vb{k}} y}^{(1)}\partial_t s_{{\vb{k}} x}^{(1)}}_{t}
\nline
&=\abs{A_0}^2\frac{v_{kx}^2\xi_{\vb{k}}\Delta}{E_{\vb{k}}^3}\frac{2\gamma\omega^2E_{\vb{k}}}{(4E_{\vb{k}}^2-\omega^2)^2+(4\gamma\omega E_{\vb{k}})^2}
\,.
\end{align}
In the small damping limit ($\gamma \rightarrow 0$), the last term in \equa{eqn:sy2} becomes a delta function corresponding to the Fermi golden rule of resonant transition.
Summing over all the pseudo-spins, $F_\theta
=2\Delta\sum_{\vb{k}} \expval*{s_{{\vb{k}} y}^{(2)}}$, one obtains \equa{ei_force}.
A similar calculation for $\expval*{s_{{\vb{k}} x}^{(2)}}$ (with the $\vb{B}_{\vb{k}}^{(2)}$ term included) yields the force on the amplitude derived in Appendix~D of Ref.~\cite{sun2024floquet}.

Note that the response of the order parameter also feeds back to the total pseudomagnetic field $\vb{B}_{\vb{k}}^{\rm tot}$.
In an ideal s-wave excitonic insulator without $p$-wave pairing channels,
the order parameter fluctuations do not respond to the uniform optical field at linear order~\cite{Sun_BaSh_2020}.
Therefore, these effects are absent in our results for the effective force at order $A_0^2$.

\setcounter{equation}{0}
\renewcommand{\theequation}{C\arabic{equation}}
\renewcommand{\theHequation}{C.\arabic{equation}}
\emph{Appendix C: Symmetry and topology constraints on the effective force---}If the slow degrees of freedom form a closed manifold $M$ where any two points are related by a symmetry of the driven system, a nonzero effective force must be nonconservative.
If it were conservative, it would be a gradient (exterior derivative) of a potential, which by symmetry has to be a constant on $M$, meaning that the force must be zero.
The NCSFs on the phase variable in the EI and CDW systems are the typical examples where $M$ has the topology of $S^1$.
The same force structure exists in a wide class of examples such as a single particle in a propagating wave and a biased Josephson junction, which may be called the ``photon-drag'' class, see SI Sec.~I~\cite{supp} for details.

Secondly, the symmetry may be high enough to make the nonconservative force field vanish too.
For example, the slow degrees of freedom of an SU(2) ferromagnet form a sphere $S^2$.
A point in $S^2$, e.g., the north pole, is invariant under rotation around the $z$ axis, while the rotation acts non-trivially on the force $F$ here.
Note that as a natural $1$-form on $M$, the force $F$ transforms under a symmetry operation following the pullback map of the symmetry operation on $M$. 
Since symmetry requires the force field to be invariant under the nontrivial transformation, it must be zero.
Therefore, any drive of the ferromagnet that respects the  SU(2) symmetry  cannot induce a static force on its angular direction.

\ifarXiv
    \onecolumngrid
    \newpaxsetup{addannots=true}
    

\section*{Supplementary material}

# Supplemental Material for 'Light-induced nonconservative static forces in many-body systems'

Jinze Li[1,*], Xuemei Yang[1,*], and Zhiyuan Sun[1,2,†]
[1]State Key Laboratory of Low-Dimensional Quantum Physics and Department of Physics, Tsinghua University, Beijing 100084, P. R. China
[2]Frontier Science Center for Quantum Information, Beijing 100084, P. R. China


CONTENTS



## I. PHOTON-DRAG CLASS OF THE NCSF

A very simple example of an NCSF arises in the following equation:

$$m(\ddot{x} + \gamma\dot{x}) = E_0 \sin(qx - \omega t) \tag{S1}$$

which describes the motion of a one-dimensional charged particle in a propagating electric field: $x$ is its coordinate, $m$ is its mass, and $\gamma$ is the damping rate (friction coefficient). We solve the motion of $x = x_0 + x^{(1)} + x^{(2)} + O(E_0^3)$ perturbatively in the driving field $E_0$. The linear response is

$$x^{(1)}(t) = \frac{-E_0}{m(\omega^2+\gamma^2)}\left[\sin\left(qx_0 - \omega t\right) - \frac{\gamma}{\omega}\cos\left(qx_0 - \omega t\right)\right]. \tag{S2}$$

Plugging $x_0 + x^{(1)}$ back into Eq. (S1) and time-averaging over one period, one obtains a static drag force on $x^{(2)}$:

$$m\left(\ddot{x}^{(2)} + \gamma\dot{x}^{(2)}\right) = F_{\mathrm{dc}}^{(2)},$$
$$F_{\mathrm{dc}}^{(2)} = qE_0\left\langle x^{(1)}\cos\left(qx_0 - \omega t\right)\right\rangle_t = \frac{qE_0^2}{2m}\frac{\gamma}{\omega(\omega^2+\gamma^2)}. \tag{S3}$$

This rectified static force could be viewed as acting on the slow component of $x$. It points to the direction of propagation of the wave, and is thus understood as photons dragging the particle. If the system lives on a ring, it is unambiguous that this drag force is nonconservative. Apparently, the nonconservative force is proportional to damping $\gamma$. In an electron gas, this effect manifests as a photo-drag DC current.

In addition to the photon-drag example, the same equation describes many physical systems. The three representative realizations are summarized in Table I.

TABLE I. Representative examples in the photon-drag class of nonconservative effective static forces.

| System | Dynamical equation | NCSF | Measurable quantity |
|---|---|---|---|
| Single-particle photon drag | $m(\ddot{x}+\gamma\dot{x}) = eE_0\sin(qx-\omega t)$ | $F^{(2)}_{\text{drag}} = \frac{e^2E_0^2q}{2m}\frac{\gamma}{\omega(\omega^2+\gamma^2)}$ | DC drift velocity; photo-drag DC current. |
| Biased Josephson junction | $\ddot{\varphi}+\gamma\dot{\varphi} = -\omega_J^2\sin(\varphi+\mu t)$ | $F^{(2)}_J = -\frac{\omega_J^4}{2}\frac{\gamma}{\mu(\mu^2+\gamma^2)}$ | Correction to the phase winding velocity, oscillating Josephson current or the emitted radiation. |
| In-plane electric dipole under circularly polarized light | $I\big(\ddot{\theta}+\gamma\dot{\theta}\big) = E_0P_0\sin(\theta-h\omega t)$ | $F^{(2)}_\theta = \frac{h(E_0P_0)^2}{2I}\frac{\gamma}{\omega(\omega^2+\gamma^2)}$ | Angular drift and electromagnetic radiation from the rotating dipole. |

The second example is the AC Josephson effect regime of a biased Josephson junction (voltage biasing a Josephson junction and a resistor in series, or current biasing a Josephson junction and a resistor in parallel). The equation of motion is $\ddot{\varphi}+\gamma(\dot{\varphi}-\mu)+\omega_J^2\sin\varphi = 0$ where $\mu = 2eV/\hbar$ and $V$ is the constant bias voltage and $\omega_J$ is the Josephson plasma frequency. Changing to the 'rotating frame' by the change of variable $\varphi = \mu t+\theta$, one obtains $\ddot{\theta}+\gamma\dot{\theta}+\omega_J^2\sin(\theta+\mu t) = 0$, which is the same as Eq. (S1). There is a second-order static drag force $F^{(2)}_J = -\omega_J^4\gamma/\big[2\mu(\mu^2+\gamma^2)\big]$ on the slow component of $\theta$. This NCSF renders a constant drift of $\theta$ with the velocity $-\omega_J^4/\big[2\mu(\mu^2+\gamma^2)\big]$, which constitutes a negative correction to the winding frequency of the phase $\varphi$. This effect could be measured by the frequency of the oscillating Josephson current or the emitted radiation.

The third example is an in-plane dipole $\mathbf{P} = P_0(\cos\theta,\sin\theta)$ driven by circularly polarized light. This applies to dipolar molecules and to the intravalley exciton condensates with an in-plane polarization associated with its phase [1, 2]. The equation of motion for its angle is given in Table I. Here $I$ is the rotational moment of inertia and $h = \pm1$ denotes the optical helicity. The NCSF induces an angular drift that can be detected through its electromagnetic radiation. Note that the frequency of the radiation is set by the angular winding rate, which is not equal to the frequency of the driving light field.

This collection of examples is thus called the 'photon-drag class' of the NCSF.

## II. NONLINEAR PHONONICS

This section contains detailed calculations and realistic parameters for the nonlinear phononics example. We consider a fast infrared-active phonon $X(t)$ coupled to slow coordinates $\phi_1(t)$ and $\phi_2(t)$ through the Lagrangian in Eq. (2) of the main text. To include the damping rate $\gamma$, we employ the Keldysh Lagrangian:

$$\begin{aligned} L = \frac{1}{2}\big(X_{\text{cl}}\;\; X_{\text{q}}\big)&\begin{pmatrix} \omega_{\text{IR}}^2\phi_{1,\text{q}} & -\partial_t^2+\gamma\partial_t-(1-\phi_{1,\text{cl}})\omega_{\text{IR}}^2 \\ -\partial_t^2-\gamma\partial_t-(1-\phi_{1,\text{cl}})\omega_{\text{IR}}^2 & \frac{1}{4}\omega_{\text{IR}}^2\phi_{1,\text{q}}-2\gamma\partial_t\coth\frac{i\partial_t}{2T}\end{pmatrix}\begin{pmatrix}X_{\text{cl}}\\ X_{\text{q}}\end{pmatrix} \\ &+E(t)\big(\phi_{2,\text{q}}\;\;\phi_{2,\text{cl}}\big)\begin{pmatrix}X_{\text{cl}}\\ X_{\text{q}}\end{pmatrix}+L_s[\phi_{1\text{cl/q}},\phi_{2\text{cl/q}}]\,. \end{aligned} \tag{S4}$$

In the following, the quantum and thermal fluctuations contained in the Keldysh kernel $\gamma\partial_t\coth\frac{i\partial_t}{2T}$ will be neglected since they do not affect the driving-field-dependent effective force [3]. Integrating out $X$ exactly and retaining terms linear in $\phi_{1\text{q}}$, $\phi_{2\text{q}}$, one obtains the drive-induced effective Lagrangian:

$$L_{\text{eff}} = \frac{1}{4}\frac{\omega_{\text{IR}}^2\phi_{1,\text{q}}\phi_{2,\text{cl}}^2+2\phi_{2,\text{q}}\phi_{2,\text{cl}}\big[\omega_{\text{IR}}^2(1-\phi_{1,\text{cl}})-\omega^2\big]}{\big[\omega_{\text{IR}}^2(1-\phi_{1,\text{cl}})-\omega^2\big]^2+\gamma^2\omega^2}|E_0|^2. \tag{S5}$$

From $F_1 = \delta L_{\text{eff}}/\delta\phi_{1,\text{q}}$ and $F_2 = \delta L_{\text{eff}}/\delta\phi_{2,\text{q}}$ and taking the classical limit $\phi_{1,\text{cl}}\to\phi_1$ and $\phi_{2,\text{cl}}\to\phi_2$, one obtains the static force in Eq. (3) of the main text.

The slow variables have a direct interpretation in terms of Raman-mode coordinates. If the physical Raman displacement $Q_1$ shifts the squared infrared-phonon frequency as $\omega_{\text{IR}}^2\to\omega_{\text{IR}}^2-aQ_1$, then $\phi_1 = aQ_1/\omega_{\text{IR}}^2$ and $a$ is the anharmonic coupling constant. The second coordinate parametrizes the displacement dependence of the infrared-mode effective charge, $\phi_2 = Z^*(Q_2) = Z_0^*+\lambda Q_2$, where $\lambda = \partial Z^*/\partial Q_2$ is the light coupling constant. When $Z^*$ is expressed in units of $e$, $\phi_2$ is dimensionless. More generally, one may normalize $Z^*$ by a reference charge and absorb that scale into the definition of the electric field. For the perovskite oxide $PrMnO_3$ [4] used in Fig. 1 of the main text, the driven $B_{1u}$ infrared mode has the intrinsic frequency $\omega_{\text{IR}} = 622\ \text{cm}^{-1}$ and damping rate $\gamma = 0.05\omega_{\text{IR}}$ while the drive frequency is $\omega = 600\ \text{cm}^{-1}$. The two $A_g$ Raman modes associated with $Q_1$ and $Q_2$ have frequencies $\omega_1 = 155\ \text{cm}^{-1}$ and $\omega_2 = 97\ \text{cm}^{-1}$, respectively, and the cubic anharmonic coupling is $a = 51.74\ \text{meV}/(\text{amu}\cdot\text{Å}^3)$.

The equilibrium restoring force of the Raman coordinate $\phi_1$ adds the contribution $-k_R\phi_1$ to $F_1$. Under the optical field, if the damping of the mode is small, the Lorentzian peak structure of $F_1$ would lead to multiple fixed points [5]. For the damping rate we use, the Lorentzian peak is broad and there is only one fixed point $(\phi_{1,0}, \phi_{2,0})$ determined by

$$\phi_{1,0} = 1 - \frac{\omega^2}{\omega_{\mathrm{IR}}^2}, \quad \phi_{2,0} = \left(\frac{4\gamma^2\omega^2 k_R \phi_{1,0}}{\omega_{\mathrm{IR}}^2 e^2 |E_0|^2}\right)^{1/2}. \tag{S6}$$

A restoring force along $\phi_2$ has been neglected since it shifts the fixed point quantitatively without changing the existence of the nonconservative circulation. If the NCSF is strong enough, the resulting dynamics can support limit cycles around this point. Since it is not the focus of the current work, this intriguing behavior is left for future research.

## III. DRIVEN EXCITONIC INSULATOR

This section contains the derivation of the Keldysh effective action for the order parameter of an EI. The effective forces are contained in this effective action. The formal structure is useful for describing the collective dynamics of other ordered states such as charge-density waves. Starting from Eq. (4) of the main text, we write the excitonic field as $(\Delta + \delta)e^{i\theta}$, where $\delta$ and $\theta$ denote amplitude and phase fluctuations, respectively. Expanding the Lagrangian to quadratic order in the fluctuations, neglecting spatial gradients, yields

$$L[\Psi, A, \delta, \theta] = \bar{\Psi}\left[-i\partial_t + \xi_{\mathbf{p}}\sigma_3 + \Delta\sigma_1\right]\Psi + V[\Psi, A, \delta, \theta] + \frac{1}{g}(\Delta + \delta)^2, \tag{S7}$$

with

$$V[\Psi, A, \delta, \theta] = \bar{\Psi}\left[\sigma_3\left(\mathbf{v_p}\cdot\mathbf{A} + \frac{1}{2m}A^2\right) + \sigma_1\left(\delta - \frac{1}{2}\Delta\theta^2\right) - \sigma_2(\Delta+\delta)\theta\right]\Psi\,. \tag{S8}$$

Here $\sigma_i$ are Pauli matrices acting in the band pseudospin space. We set the Planck constant $\hbar$, speed of light $c$, and electron charge $e$ to unity for notational simplicity but will occasionally restore them for physical results. The three terms in Eq. (S7) correspond to the free quasiparticle sector, the fluctuation and electromagnetic coupling sector, and the Hubbard-Stratonovich sector associated with excitonic pairing. Note that the $\mathbf{v_k} = \partial_{\mathbf{k}}\xi_{\mathbf{k}}$ in Eq. (S8) is the group velocity in the normal state with $\Delta = 0$.

Correspondingly, the Keldysh Lagrangian is $L[\Psi, A_{\mathrm{cl/q}}, \delta_{\mathrm{cl/q}}, \theta_{\mathrm{cl/q}}] = L_+ - L_-$, the difference between the Lagrangians on the forward and backward branches of the Keldysh contour, following the standard notation for fermionic and bosonic fields [6]. The interaction vertex is now written as $\bar{\Psi}\hat{V}\Psi$ where $\hat{V} = V_{\mathrm{cl}}\tau_0 + \frac{1}{2}V_{\mathrm{q}}\tau_1$ and $\tau_0$ and $\tau_1$ are the usual Pauli matrices acting in Keldysh space. Explicitly,

$$V_{\mathrm{cl}} = \sigma_3\left(\mathbf{v_p}\cdot\mathbf{A} + \frac{1}{2m}A^2\right) + \sigma_1\left(\delta_{\mathrm{cl}} - \frac{1}{2}\Delta\theta_{\mathrm{cl}}^2\right) - \sigma_2(\Delta+\delta_{\mathrm{cl}})\theta_{\mathrm{cl}}\,, \tag{S9a}$$

$$V_{\mathrm{q}} = \sigma_1\left(\delta_{\mathrm{q}} - \Delta\theta_{\mathrm{q}}\theta_{\mathrm{cl}}\right) - \sigma_2\left[\theta_{\mathrm{q}}(\Delta+\delta_{\mathrm{cl}}) + \delta_{\mathrm{q}}\theta_{\mathrm{cl}}\right]\,. \tag{S9b}$$

Note that the external electromagnetic field is a classical drive field so that only a classical component is retained. For the collective fields $\theta$ and $\delta$, we keep terms only to linear order in the quantum components, corresponding to the semiclassical approximation relevant for the collective dynamics considered here. The Hubbard-Stratonovich term reads $\hat{L}_{\mathrm{HS}} = \frac{2}{g}(\Delta + \delta_{\mathrm{cl}})\delta_{\mathrm{q}}$.

In the Keldysh basis [6], the free quasiparticle Green function defined from the mean-field Hamiltonian reads

$$\hat{G}_0 = -i\langle\Psi\bar{\Psi}\rangle_0 = \begin{pmatrix} G_0^R & G_0^K \\ 0 & G_0^A \end{pmatrix},$$

where

$$G_0^{R/A}(\mathbf{k},\epsilon) = \frac{(\epsilon \pm i\eta) - \xi_{\mathbf{k}}\sigma_3 - \Delta\sigma_1}{(\epsilon \pm i\eta - E_{\mathbf{k}})(\epsilon \pm i\eta + E_{\mathbf{k}})}, \tag{S10a}$$

$$G_0^K(\mathbf{k},\epsilon) = \left[G_0^R(\mathbf{k},\epsilon) - G_0^A(\mathbf{k},\epsilon)\right]\tanh\frac{\epsilon}{2T}, \tag{S10b}$$

and $E_{\mathbf{k}} = \sqrt{\xi_{\mathbf{k}}^2 + \Delta^2}$ is the positive quasiparticle energy.

### A. Correlators

Integrating out the fermions yields the effective action in powers of the interaction vertex, $S_{\rm eff} = \sum_n S_{\rm eff}^{(n)} + S_{\rm HS} = \sum_n i\frac{(-1)^n}{n}\,{\rm tr}\left[(\hat{G}_0\hat{V})^n\right] + S_{\rm HS}$, where ${\rm tr}[\cdots] \equiv \int dt d\mathbf{k}\,{\rm tr}_{\sigma,\tau}[\cdots]$ includes both momentum integration and trace over band/Keldysh space. The first-order term is particularly simple, which reads $S_{\rm eff}^{(1)} = -(\delta_{\rm q} - \Delta\theta_{\rm q}\theta_{\rm cl})\sum_{\mathbf{k}} \frac{\Delta}{E_{\mathbf{k}}}\tanh\frac{E_{\mathbf{k}}}{2T}$ at first order in the quantum component of the field. The saddle-point equation $\partial_{\delta_{\rm q}} S_{\rm eff} = 0$ for the amplitude fluctuation reads $\frac{2}{g}\Delta = \sum_{\mathbf{p}} \frac{\Delta}{E_{\mathbf{p}}}\tanh\frac{E_{\mathbf{p}}}{2T}$, which is the gap equation determining the equilibrium mean-field order parameter $\Delta$.

To group the higher-order terms, it is convenient to introduce two retarded correlation functions. The retarded two-point correlation function is

$$\begin{aligned}\chi_{\Gamma_1\Gamma_2}(\omega) &= -i\int dt e^{-i\omega t}\Theta(-t)\left\langle\left[\hat{O}_2(0),\hat{O}_1(t)\right]\right\rangle = -i\int dt e^{i\omega t}\left\langle O_{2,{\rm cl}}(0)O_{1,{\rm q}}(t)\right\rangle \\ &= \frac{i}{2}\sum_{\mathbf{k},\Omega}{\rm tr}_\sigma\left[G_0^R(\omega+\Omega)\Gamma_1 G_0^K(\Omega)\Gamma_2 + G_0^K(\omega+\Omega)\Gamma_1 G_0^A(\Omega)\Gamma_2\right],\end{aligned} \tag{S11}$$

corresponding to the ‘bubble’ diagram. Here $\hat{O}_i = \sum_{\mathbf{k}} \Psi_{\mathbf{k}}^\dagger\Gamma_i\Psi_{\mathbf{k}}$ and $\Gamma_i$ denotes the corresponding vertex and $\Theta(t)$ is the Heaviside step function. In Eq. (S11), the first $\langle\rangle$ means quantum mechanical expectation value over the density matrix, while the second $\langle\rangle$ means the Keldysh path integral average. The momentum indices of the Green functions are omitted for notational simplicity. The retarded three-point correlation function is

$$\begin{aligned}\chi_{\Gamma_1\Gamma_2\Gamma_3}(\omega+\omega',-\omega;\omega') &= (-i)^2\int dt_1 dt_2 e^{-i(\omega+\omega')t_1+i\omega t_2}\Theta(-t_1)\Theta(-t_2)\left\langle\left[\left[\hat{O}_3(0),\hat{O}_2(t_2)\right],\hat{O}_1(t_1)\right]\right\rangle \\ &= (-i)^2\int dt_1 dt_2 e^{-i(\omega+\omega')t_1+i\omega t_2}\left\langle O_{3,{\rm cl}}(0)O_{2,{\rm q}}(t_2)O_{1,{\rm q}}(t_1)\right\rangle \\ &= -\frac{i}{2}\sum_{\mathbf{k},\Omega}{\rm tr}_\sigma\begin{pmatrix} G_0^K(\Omega)\Gamma_1 G_0^A(\Omega+\omega+\omega')\Gamma_2 G_0^A(\Omega+\omega')\Gamma_3 \\ +G_0^R(\Omega)\Gamma_1 G_0^K(\Omega+\omega+\omega')\Gamma_2 G_0^A(\Omega+\omega')\Gamma_3 \\ +G_0^R(\Omega)\Gamma_1 G_0^R(\Omega+\omega+\omega')\Gamma_2 G_0^K(\Omega+\omega')\Gamma_3\end{pmatrix},\end{aligned} \tag{S12}$$

corresponding to the ‘triangular’ diagram in Fig. 2(a) of the main text. For the rectified dc response generated under optical driving, one should set the output frequency $\omega' \to 0$. In some cases, the $\chi$ diverges in this limit, and one may need to analyse this divergence occasionally. Note that the frequency integral over $\Omega$ in Eq. (S12) may discourage the contour integral technique because of the $\tanh\frac{\epsilon}{2T}$ that encodes the occupation number in the Keldysh Green function $G_0^K(\epsilon)$. Nevertheless, for optical pulses where the typical $\omega'$ is larger than the system-bath relaxation rate and inter-particle thermalization rate, this occupation factor could be taken as $\tanh\frac{E_{\mathbf{k}}}{2T}$ as it only matters for the sharp Lorentzian on the real frequency axis at $\epsilon = E_{\mathbf{k}}$ [5]. With this substitution, contour integral can be used.

### B. Effective action of the phase and amplitude modes

We integrate out the fermions to obtain the effective action $S_{\rm eff} = \int dt L_{\rm eff}[\Delta, A]$ for the order parameter, i.e., that of the phase and amplitude modes. The effective Keldysh Lagrangian is obtained as

$$L_{\rm eff} = L_\theta[\theta,A] + L_\delta[\delta,A] + L_I[\delta,\theta]\,,\quad L_\theta = \theta_{\rm q}G_\theta^{-1}\theta_{\rm cl} - \theta_{\rm q}\left[F_\theta(t) + F_0(t)\right], \tag{S13}$$

$$L_\delta = \delta_{\rm q}G_\delta^{-1}\delta_{\rm cl} - \delta_{\rm q}F_\delta(t),\quad L_I = C_0\left(\theta_{\rm q}\partial_t\delta_{\rm cl} - \delta_{\rm q}\partial_t\theta_{\rm cl}\right). \tag{S14}$$

The $L_\theta$, $L_\delta$, $L_I$ are the Lagrangians for the phase-mode, amplitude mode and their linear coupling, respectively. The inverse propagators of the phase and amplitude modes are computed from the retarded two-point correlators in Eq. (S11) as:

$$G_\theta^{-1}(\omega) = \Delta^2\left(\frac{2}{g} + \chi_{\sigma_2\sigma_2}(\omega)\right) = -\omega^2\sum_{\mathbf{k}}\frac{\Delta^2}{E_{\mathbf{k}}(4E_{\mathbf{k}}^2-\omega^2)}\tanh\frac{E_{\mathbf{k}}}{2T}\,, \tag{S15}$$

$$G_\delta^{-1}(\omega) = \frac{2}{g} + \chi_{\sigma_1\sigma_1}(\omega) \simeq \left(4\Delta^2-\omega^2\right)\sum_{\mathbf{k}}\frac{1}{E_{\mathbf{k}}(4E_{\mathbf{k}}^2-\omega^2)}\tanh\frac{E_{\mathbf{k}}}{2T}\,. \tag{S16}$$

Note that in time domain, these propagators are nonlocal in general and the terms like $\theta_{\rm q}G_\theta^{-1}\theta_{\rm cl}$ should be understood as $\theta_{\rm q}(t)\int dt' G_\theta^{-1}(t,t')\theta_{\rm cl}(t')$. In the low frequency limit, the phase kernel reduces to $G_\theta^{-1} = \frac{1}{4}\nu\partial_t^2$, consistent with the gapless

nature of the phase mode. The amplitude kernel $G_\delta^{-1}(\omega)$ exhibits the characteristic resonance structure of the Higgs-like amplitude mode near $\omega \simeq 2\Delta$. In the low frequency limit (slow limit), it reduces to $G_\delta^{-1}(\omega) = \nu\left(-\frac{\omega^2}{4\Delta^2}+1\right)$. At zero-temperature, the inertia $\nu$ is simply the normal state electronic density of states $\nu_0 = m/(\pi)$ at the Fermi level (counting two bands) before gap opening in the BCS regime ($E_{\rm G} < 0, \Delta \ll |E_{\rm G}|$). In the BEC limit ($E_{\rm G} > 0$, $\Delta \ll E_{\rm G}$), it becomes $\nu \approx \nu_0\Delta^2/\tilde{E}_{\rm G}^2$ where $\tilde{E}_{\rm G} = \sqrt{4\Delta^2 + E_{\rm G}^2}$ is the optical gap in the BEC case.

The phase-amplitude coupling coefficient is

$$C_0 = \frac{1}{i\omega}\Delta\chi_{\sigma_1\sigma_2}(\omega) \xrightarrow{\omega\to 0} \sum_{\mathbf{k}} \frac{\xi_{\mathbf{k}}\Delta}{2E_{\mathbf{k}}^3}\tanh\frac{E_{\mathbf{k}}}{2T}, \tag{S17}$$

which reduces to a constant in the low-frequency limit. This coupling vanishes in the BCS weak coupling limit because of 'electron-hole' symmetry. However, it exists in general and would affect the light-induced dynamics of the phase. In the BEC limit, it goes to $C_0 = \nu_0\Delta/(2\tilde{E}_{\rm G})$.

At second order in the optical driving field, light generates effective forces on the phase variable, which contain two physically distinct contributions. The first contribution is the NCSF derived from the triangular diagram in Fig. 2(a) in the main text (Eq. (S12)):

$$\begin{aligned} F_\theta &= \frac{1}{2}|A_0|^2\Delta\chi^{(2)}_{v\sigma_3,v\sigma_3,\sigma_2}(\omega,-\omega;0) = \frac{1}{2}|A_0|^2\sum_{\mathbf{k}}\pi\frac{v_{px}^2\xi_{\mathbf{k}}\Delta^2}{E_{\mathbf{k}}^3}\delta(\omega-2E_{\mathbf{k}})\tanh\frac{E_{\mathbf{k}}}{2T} \\ &= \frac{e^2|E_0|^2}{4\Delta}\begin{cases} \dfrac{2\Delta^3\sqrt{\omega^2-4\Delta^2}}{\omega^4}\Theta(\omega-2\Delta), & \text{BCS limit,} \\ \dfrac{\Delta^3(\sqrt{\omega^2-4\Delta^2}-E_G)}{\omega^4}\Theta\big(\omega-\tilde{E}_{\rm G}\big), & \text{BEC limit.}\end{cases} \end{aligned} \tag{S18}$$

It originates from resonant interband transitions and represents a genuine nonconservative drag force acting on the phase mode. This term remains finite under continuous irradiation and therefore produces a nonzero net impulse during the pulse duration. The second contribution,

$$F_0 = \left(\partial_t|A_0|^2\right)\Delta\lim_{\omega'\to 0}\frac{1}{i\omega'}\left(\frac{1}{4m}\chi_{\sigma_3\sigma_2}(\omega') + \frac{1}{2}\chi^{(2)}_{v\sigma_3,v\sigma_3,\sigma_2}(\omega+\omega',-\omega;\omega')\right) \tag{S19}$$

is from non-resonant processes where $\omega'$ is the frequency of the force leg. It may be understood as the generalized ponderomotive shift of the energy of the exciton condensate. Note that this force is proportional to the temporal variation of the optical intensity envelope ($F_0 \propto \partial_t A_0^2$) and would be zero for continuous wave (CW) drive. Therefore, it does not belong to the static effective force in a periodically driven system. For a light pulse, this force changes sign between the rising and falling edges of the pulse and yields zero net impulse $\int_{\rm pulse} dt F_0(t) = 0$. However, the pulse could still cause a constant shift of the phase via this effect.

The second-order effective force on the amplitude is [3]

$$F_\delta = -|A_0|^2\left[\frac{1}{4m}\chi_{\sigma_3\sigma_1}(0) + \frac{1}{2}\lim_{\omega'\to 0}\chi^{(2)}_{v\sigma_3,v\sigma_3,\sigma_1}(\omega+\omega',-\omega;\omega')\right]. \tag{S20}$$

Under CW drive, $F_\delta$ has a static component $\propto A_0^2$ and also a component that grows in time: $\propto \int dt A_0^2$. This growing component can be separated from the second term of Eq. (S20) that diverges as $1/\omega'$ in the static limit:

$$\begin{aligned} \partial_t F_\delta &= -\lim_{\omega'\to 0}\frac{1}{2}|A_0|^2(-i\omega')\chi^{(2)}_{v\sigma_3,v\sigma_3,\sigma_1}(\omega+\omega',-\omega;\omega') = -|A_0|^2\sum_{\mathbf{k}}\pi\frac{v_{px}^2\Delta^3}{E_{\mathbf{k}}^3}\delta(\omega-2E_{\mathbf{k}})\tanh\frac{E_{\mathbf{k}}}{2T}. \\ &= -\frac{e^2|E_0|^2}{\Delta}\begin{cases} \dfrac{2\Delta^4|E_G|}{\omega^4\sqrt{\omega^2-4\Delta^2}}\Theta(\omega-2\Delta), & \text{BCS limit,} \\ \dfrac{\Delta^4(\sqrt{\omega^2-4\Delta^2}-E_G)}{\omega^4\sqrt{\omega^2-4\Delta^2}}\Theta\big(\omega-\tilde{E}_{\rm G}\big), & \text{BEC limit.}\end{cases} \end{aligned} \tag{S21}$$

Physically, this force grows because the optically excited quasiparticles grow linearly in time, which tends to reduce the gap. With a bath that takes away the heat with rate $\gamma_{\rm E}$, this force becomes $\sim 1/\gamma_{\rm E}$ in the steady state, see Eq. D6 of Appendix D of Ref. [3].

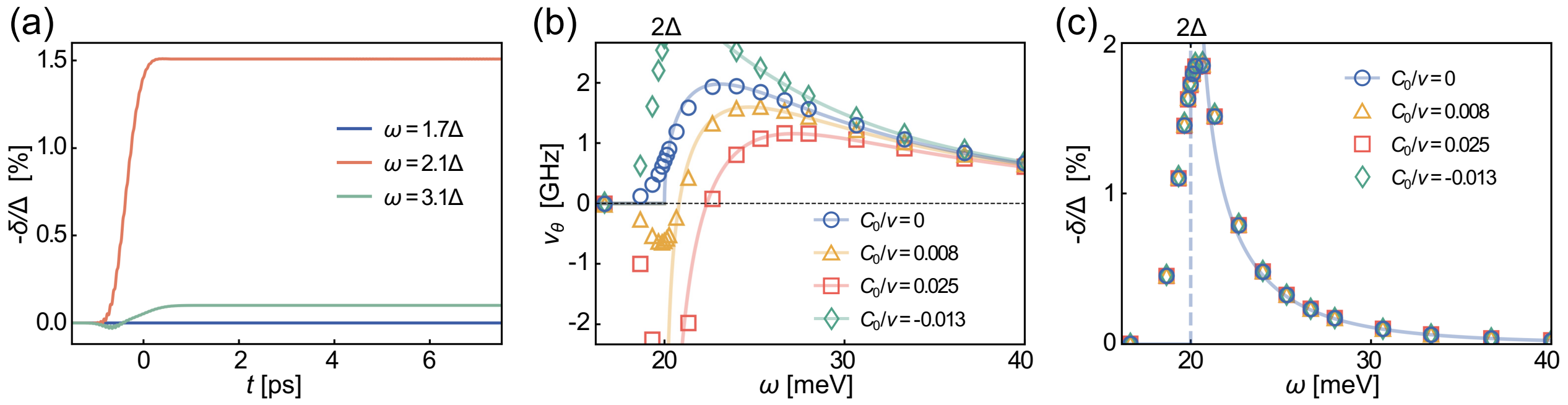


FIG. S1. Simulated dynamics of a BCS-type excitonic insulator driven by a Gaussian optical pulse (centered at time zero) with duration $\tau = 0.5$ ps and peak electric field $E_0 = 250$ V/cm. (a) Time dependence of the order parameter amplitude for different central frequencies of the pulse. (b) Light-frequency dependence of the resulting phase velocity (after the pulse) for different phase-amplitude coupling strengths $C_0$. Markers and lines represent numerical and analytical results, respectively, with $C_0$ controlled by the ultraviolet cutoff. (c) Light-frequency dependence of the amplitude-mode displacement, showing weak dependence on $C_0$. The deviations near the gap in (b) and (c) from analytical results originate from higher-order effects in the driving field, which are most pronounced close to the optical gap. The parameters are $E_G = -100$ meV, $\Delta = 10$ meV and $m = 0.2m_e$.

### C. Slow equation of motion

In the slow limit (dynamics much slower than the light frequency $\omega$ and the gap $2\Delta$), the effective Lagrangian is written as

$$L_{\rm eff}[\theta, \delta, A] = \theta_{\rm q}\left(\frac{1}{4}\nu\partial_t^2\theta_{\rm cl} - C_0\partial_t\delta_{\rm cl} - F_\theta - F_0\right) + \delta_{\rm q}\left[\nu\left(\frac{1}{4\Delta^2}\partial_t^2 + 1\right)\delta_{\rm cl} + C_0\partial_t\theta_{\rm cl} - F_\delta\right], \tag{S22}$$

which gives Eq. (5) in the main text after noting that $F_0$ is not a static force. Eliminating the amplitude fluctuation $\delta_{\rm cl}$ gives an effective equation of motion for the phase:

$$\frac{1}{4}\tilde{\nu}\partial_t^2\theta_{\rm cl} = F_\theta + F_0 + \frac{C_0}{\nu}\partial_t F_\delta \tag{S23}$$

which is Eq. (6) in the main text. Note that in the BCS weak-coupling case in 2D with a symmetric UV cutoff relative to the Fermi level, "particle-hole" symmetry becomes exact for a parabolic band, and the amplitude-phase coupling $C_0$ vanishes exactly. In this case, the equation of motion for the phase simplifies to $\frac{\nu}{4}\partial_t^2\theta = F_\theta + F_0$.

For a pulsed pump with width $\tau$, the dc component of the effective phase force has contributions only from $F_\theta$ and $\partial_t F_\delta$, both of which are proportional to the absorbed power. The resulting phase velocity is estimated from Eq. (S23) as $v_\theta = \sqrt{\pi}\tau\frac{4}{\tilde{\nu}}\left(F_\theta + \frac{C_0}{\nu}\partial_t F_\delta\right)$. With a nonzero amplitude-phase coupling $C_0$, the amplitude force $\partial_t F_\delta$ contributes to the acceleration of the phase. From Eq. (S22), the contribution of the amplitude force to the terminal phase velocity is proportional to the static amplitude displacement after the pulse. Physically, the nonthermal 'heating' from the light pulse reduces the amplitude of the order parameter after the pulse, which leads to an additional phase winding through the amplitude-phase coupling. These features are confirmed by direct numerical simulations of the mean-field dynamics, see Fig. S1.

### D. BEC limit

In the dilute BEC case ($|\omega_0| \ll E_G$), the phase-amplitude coupling $C_0$ dominates and the inertia terms are not important. As a result, the low-energy effective Lagrangian crosses over to the Gross-Pitaevskii-type for the excitonic field $\Phi$: $L[\Phi] = \Phi^*(-i\partial_t + \omega_0)\Phi + g_0|\Phi|^4$, where $g_0$ is the exciton-exciton interaction. Excitonic BEC with the density $\rho_0 = |\Phi|^2 = -\omega_0/(2g_0)$ occurs when $\omega_0 < 0$. In this regime, the phase $\theta$ and amplitude $\rho$ of the order parameter $\Phi = \sqrt{\rho}e^{i\theta}$ form a pair of conjugate variables. The equation of motion implied by the Gross-Pitaevskii Lagrangian reads

$$\dot{\theta} = -\omega_0 - 2g_0\rho + F_\rho, \quad \dot{\rho} = -F_\theta, \tag{S24}$$

where we have added the light-induced effective static forces $F_\rho$ and $F_\theta$ acting on $\rho$ and $\theta$, respectively. Eliminating the amplitude variable, one obtains the equation of motion for the phase:

$$\ddot{\theta} = 2g_0F_\theta + \dot{F}_\rho. \tag{S25}$$

This equation can also be obtained from Eq. (5) in the main text in the limit $\nu\partial_t^2 \to 0$, since in the BEC regime the phase inertia stiffness is negligible and the dynamics is governed primarily by the canonical amplitude–phase structure.

### 1. BEC-type excitonic insulator from electron-hole Coulomb interaction

In realistic two-dimensional materials, the nonlocality of long-range Coulomb interactions can significantly affect both phase and amplitude dynamics. To capture this effect, we consider a model with interlayer electron–hole Coulomb interaction $V_{\mathbf{k},\mathbf{k}'} = 2\pi e^2/(\epsilon|\mathbf{k}-\mathbf{k}'|)$, where $\epsilon$ is the effective dielectric constant of the environment. In this case, the mean-field order parameter acquires momentum dependence, reflecting the internal structure of the exciton bound state $\Delta_{\mathbf{k}} = -\sum_{\mathbf{k}'} V_{\mathbf{k},\mathbf{k}'}\langle\psi^\dagger_{1,\mathbf{k}'}\psi_{2,\mathbf{k}'}\rangle$. With an interlayer chemical potential bias $\mu$, we work in the rotating frame where the bare band gap is already reduced from $G$ to $E_G = G-\mu$. The Lagrangian is given by [1]

$$L[\Psi,\Delta,\mathbf{A}] = \sum_{\mathbf{k}} \Psi^\dagger \begin{pmatrix} -i\partial_t + \xi_{\mathbf{k}+\mathbf{A}} & \Delta_{\mathbf{k}} \\ \Delta^*_{\mathbf{k}} & -i\partial_t - \xi_{\mathbf{k}+\mathbf{A}} \end{pmatrix} \Psi + \sum_{\mathbf{k},\mathbf{k}'} (V^{-1})_{\mathbf{k},\mathbf{k}'} \Delta^*_{\mathbf{k}}\Delta_{\mathbf{k}'}. \tag{S26}$$

Here $\xi_{\mathbf{k}} = k^2/(2m) + E_G/2$. For a generic attractive interaction, the pairing field can be expanded in the eigenbasis of the two-body Bethe–Salpeter equation. Keeping only the lowest-energy $s$-exciton, the order parameter takes the single-mode form

$$\Delta_{\mathbf{k}}(t) = (\omega_0 - 2\xi_{\mathbf{k}})\varphi_{\mathbf{k}}\Phi(t), \tag{S27}$$

where $\varphi_{\mathbf{k}}$ is the $s$-wave bound-state wavefunction, $\omega_0$ is the bound-state energy, and $\Phi(t)$ is the bosonic field for s-excitons. For Coulomb interactions, one has

$$\varphi_{\mathbf{k}} = \sqrt{8\pi}\frac{q_0^2}{(q_0^2+k^2)^{3/2}}, \quad E_{\mathrm{b}} = \frac{me^4}{\hbar^2\epsilon^2} = \frac{\hbar^2 q_0^2}{m}, \quad 2\xi_{\mathbf{k}} - \omega_0 = \frac{\hbar^2 k^2}{m} + E_{\mathrm{b}}, \tag{S28}$$

where $q_0 = 2/a_0$ and $a_0 = 2\epsilon\hbar^2/(me^2)$ is the exciton Bohr radius.

From Eq. (S18), Eq. (S21) and Eq. (S27), the phase and amplitude forces are given by

$$F_\theta = \frac{1}{4}|eE_0|^2\Delta^2_{\mathbf{k}}\frac{mv_k^2}{\omega^4}\Theta(\omega-\tilde{E}_{\mathrm{G}}) = \frac{1}{4}|eE_0|^2\frac{8\pi\rho_0 E_{\mathrm{b}}^2/(m)}{E_{\mathrm{b}}+k^2/m}\frac{mv_k^2}{\omega^4}\Theta(\omega-\tilde{E}_{\mathrm{G}}), \tag{S29}$$

$$\partial_t F_\rho = -\frac{1}{4}|eE_0|^2\frac{\Delta^4_{\mathbf{k}}}{\omega^4}\frac{mv_k^2}{\rho_0\xi_k}\Theta(\omega-\tilde{E}_{\mathrm{G}}) = -\frac{1}{4}|eE_0|^2(8\pi)^2\frac{E_{\mathrm{b}}^4}{m^2(E_{\mathrm{b}}+k^2/m)^2}\frac{1}{\omega^4}\frac{\rho_0 mv_k^2}{\xi_k}\Theta(\omega-\tilde{E}_{\mathrm{G}}). \tag{S30}$$

Here $k$ is defined as the momentum where resonant excitation $\omega = 2E_k$ happens and $\tilde{E}_{\mathrm{G}} = \sqrt{E_G^2 + 4\Delta_0^2}$ is the optical gap. The net acceleration in Eq. (S25) is thus

$$2g_0F_\theta + \partial_t F_\rho = |eE_0|^2 4\pi\frac{E_{\mathrm{b}}^2 mv_k^2}{\omega^4}\frac{\rho_0/m}{E_{\mathrm{b}}+k^2/m}\left[g_0 - \frac{4\pi E_{\mathrm{b}}^2}{m\xi_k(E_{\mathrm{b}}+k^2/m)}\right]\Theta(\omega-\tilde{E}_{\mathrm{G}}). \tag{S31}$$

In the dilute case ($E_{\mathrm{b}} \approx E_G$), one has $\Delta_{\mathbf{k}} \ll E_{\mathrm{b}}$ and the relations $mv_k^2 = \sqrt{\omega^2 - 4\Delta^2_{\mathbf{k}}} - E_G \approx \omega - \tilde{E}_{\mathrm{G}}$, $\xi_k \approx \omega/2$ and $E_{\mathrm{b}} + k^2/m \approx \omega$. Eq. (S31) simplifies to

$$2g_0F_\theta + \partial_t F_\rho = |eE_0|^2 4\pi\frac{E_{\mathrm{b}}^2 mv_k^2}{\omega^4}\frac{\rho_0/m^2}{\omega}\left(mg_0 - \frac{8\pi E_{\mathrm{b}}^2}{\omega^2}\right)\Theta(\omega-\tilde{E}_{\mathrm{G}}). \tag{S32}$$

The acceleration is positive in the high-frequency limit, but becomes negative at low frequencies. The location of sign reversal depends on the exciton-exciton interaction $g_0$. Therefore, if the frequency of sign reversal is measured experimentally, it is also a measure of $g_0$. For simplicity, we employ the exciton-exciton interaction $g_0 \approx 4/\nu_0 = 4\pi/m$ [1] from exchange processes where $\nu_0 = m/(\pi)$. The resulting phase winding speed induced by a Gaussian light pulse $E(t) = E_0 e^{-t^2/(2\tau^2)}\cos\omega t$ of duration $\tau$ and peak field $E_0$ is therefore

$$v_\theta = \int dt\big(2g_0F_\theta + \partial_t F_\rho\big) = \sqrt{\pi}(4\pi)^2\tau|eE_0|^2\frac{\rho_0}{m^2}\frac{E_{\mathrm{b}}^2(\omega-\tilde{E}_{\mathrm{G}})}{\omega^5}\left(1 - \frac{2E_{\mathrm{b}}^2}{\omega^2}\right)\Theta(\omega-\tilde{E}_{\mathrm{G}}) \tag{S33}$$

which is Eq. (9) of the main text. The factor $(1-2E_{\mathrm{b}}^2/\omega^2)$ predicts the sign reversal frequency at $\omega = \sqrt{2}E_{\mathrm{b}}$.

### E. Numerical simulations

We numerically simulated the time-dependent self-consistent Bogoliubov–de Gennes equations for a two-band excitonic insulator with a contact interaction $g$. For each momentum $\mathbf{k}$, the driven mean-field Hamiltonian and density matrix are

$$H_{\mathbf{k}}(t) = \begin{pmatrix} \xi_{\mathbf{k}+e\mathbf{A}(t)} & \Delta(t) \\ \Delta^*(t) & -\xi_{\mathbf{k}+e\mathbf{A}(t)} \end{pmatrix}, \qquad \hat{\rho}_{\mathbf{k}}(t) = \begin{pmatrix} \langle c^\dagger_{\mathbf{k}} c_{\mathbf{k}} \rangle & \langle v^\dagger_{\mathbf{k}} c_{\mathbf{k}} \rangle \\ \langle c^\dagger_{\mathbf{k}} v_{\mathbf{k}} \rangle & \langle v^\dagger_{\mathbf{k}} v_{\mathbf{k}} \rangle \end{pmatrix} \equiv \begin{pmatrix} \rho_{11,\mathbf{k}} & \rho_{12,\mathbf{k}} \\ \rho_{21,\mathbf{k}} & \rho_{22,\mathbf{k}} \end{pmatrix}. \tag{S34}$$

The initial gap $\Delta_0$ before the optical pulse is obtained from the equilibrium gap equation, and $\hat{\rho}_{\mathbf{k}}$ is initialized in the corresponding mean-field ground state as $\hat{\rho}_{\mathbf{k}}(0) = (1 - H_{\mathbf{k}}(0)/E_{\mathbf{k}})/2$. The density matrix and the order parameter evolve according to

$$i\partial_t \hat{\rho}_{\mathbf{k}}(t) = \left[H_{\mathbf{k}}(t), \hat{\rho}_{\mathbf{k}}(t)\right], \qquad \Delta(t) = -g \sum_{\mathbf{k}} \rho_{12,\mathbf{k}}(t) \tag{S35}$$

where the second equation is the gap equation. Using hermiticity and $\mathrm{Tr}\,\hat{\rho}_{\mathbf{k}} = 1$, the evolution can be written in terms of $\rho_{11,\mathbf{k}}$ and $\rho_{12,\mathbf{k}}$ as

$$\partial_t \begin{pmatrix} \rho_{11,\mathbf{k}} \\ \rho_{12,\mathbf{k}} \end{pmatrix} = \begin{pmatrix} 2\,\mathrm{Im}\left[\Delta(t)\rho^*_{12,\mathbf{k}}(t)\right] \\ i\Delta(t)\left[2\rho_{11,\mathbf{k}}(t) - 1\right] - 2i\xi_{\mathbf{k}+e\mathbf{A}(t)}\rho_{12,\mathbf{k}}(t) \end{pmatrix}. \tag{S36}$$

We integrate these equations using a fourth-order Runge–Kutta method, updating $\Delta(t)$ self-consistently at each substep. Damping and collision terms are neglected because the pump duration is much shorter than the relevant relaxation times. Numerical convergence is verified against the momentum-grid density and time step. The numerical results in Figs. 2(bc) in the main text and Fig. S1 are from these simulations.

## IV. DRIVEN CHARGE DENSITY WAVE

For the CDW system driven by light, a similar Keldysh formalism for the EI applies but with some differences. Unlike the EI order parameter discussed above, the CDW order parameter is tied to a lattice distortion generated by the electron-phonon interaction. In one dimension for a single-band CDW, the Peierls instability selects the ordering wavevector $Q = 2k_{\mathrm{F}}$ and the order parameter is proportional to the lattice displacement. The Lagrangian reads

$$L = \Psi^\dagger \begin{pmatrix} -i\partial_t + \xi_p & \Delta \\ \Delta^* & -i\partial_t + \xi_{-p} \end{pmatrix} \Psi + \frac{1}{2g}|\Delta|^2 - \frac{|\dot{\Delta}|^2}{2g\omega_Q^2}, \tag{S37}$$

where $\xi_k = v_{\mathrm{F}} k + k^2/(2m)$ and $\Psi = (\psi_{\mathrm{R}}, \psi_{\mathrm{L}})^T$ are for right- and left-moving electrons shifted to the same point in momentum space. The off-diagonal term mixes these two species and opens a gap at the Fermi surface. The electromagnetic field enters through the minimal substitution $p \longrightarrow p + A$. For a CDW, the order parameter phase $\theta$ is not merely an internal phase variable but is also the sliding coordinate of the lattice modulation. For convenience, we make a 'local gauge transformation' of the fermion field, $\Psi = e^{i\sigma_3\theta}\Psi'$, which shifts the phase field $\theta(x,t)$ from the order parameter to the diagonal kernel. The Lagrangian expressed using the new fermion fields is $L = L_0 + V$ with

$$L_0[\Psi] = \Psi'^\dagger \left[\left(-i\partial_t + \frac{p^2}{2m}\right)\sigma_0 + v_{\mathrm{F}} p \sigma_3 + \Delta\sigma_1\right]\Psi' + \frac{1}{2g}|\Delta|^2 - \frac{|\dot{\Delta}|^2}{2g\omega_Q^2}, \tag{S38}$$

$$V[\Psi,\theta,A] = \Psi'^\dagger \left[\left(v_{\mathrm{F}}\partial_x\theta + \frac{1}{2m}\left(A^2 + (\partial_x\theta)^2 + \{p, A\}\right)\right)\sigma_0 + \left(\partial_t\theta + v_{\mathrm{F}} A + \frac{1}{m}A\partial_x\theta + \frac{1}{2m}\{p, \partial_x\theta\}\right)\sigma_3\right]\Psi'. \tag{S39}$$

The corresponding Keldysh action is obtained as in Sec. III. In the following, we also add the friction for the phase $L_{\mathrm{fric}} = \nu_{\mathrm{ion}}\gamma\theta_{\mathrm{q}}\partial_t\theta_{\mathrm{cl}}$ that can only be expressed in the Keldysh Lagrangian as it involves coupling between the forward and backward time axes. The friction for lattice motion may originate from lattice disorder and anharmonicity.

From $L_0$, the free quasiparticle Green function in the mean-field state is

$$G_0^{R/A}(k,\epsilon) = \frac{\left(\epsilon \pm i\eta - \frac{k^2}{2m}\right) - v_{\mathrm{F}} k \sigma_3 - \Delta\sigma_1}{(\epsilon \pm i\eta - E_+(k))(\epsilon \pm i\eta - E_-(k))}, \tag{S40a}$$

$$G_0^K(k,\epsilon) = \left[G_0^R(k,\epsilon) - G_0^A(k,\epsilon)\right] \tanh\frac{\epsilon}{2T}, \tag{S40b}$$

where $E_\pm(k) = \frac{k^2}{2m} \pm E_k$ are the energy dispersions of the conduction and valence bands in the CDW phase, and $E_k = \sqrt{v_F^2 k^2 + \Delta^2}$. Compared with the EI case, the quadratic curvature term breaks the exact particle-hole symmetry of the spectrum.

In the following, we integrate out the fermions to obtain the effective action $S_{\rm eff}[\theta_{\rm cl/q}, A_{\rm cl/q}] = S^{(2)}_{\rm eff} + S^{(3)}_{\rm eff} + \cdots$ for the phase and electromagnetic field.

### A. The quadratic action for the phase, linear optical response

The quadratic effective action for the CDW phase mode is obtained as

$$S^{(2)}_{\rm eff} = \int dt \Big[ \nu_{\rm ion}\theta_{\rm q}(\partial_t^2 + \gamma'\partial_t)\theta_{\rm cl} + \tilde{A}_{\rm q}\chi_{\sigma_3\sigma_3}\tilde{A}_{\rm cl} + \frac{n}{m}\left(-\theta_{\rm q}\partial_x^2\theta_{\rm cl} + A_{\rm q}A_{\rm cl}\right)\Big], \tag{S41}$$

where $n = k_F/\pi$ is the (spinless) electron density and we have defined the compact notation, the 'dressed field' $\tilde{A} = \partial_t\theta + v_F A$ that combines the phase and the vector potential. The first term comes from the last term of Eq. (S37) where $\nu_{\rm ion} = \frac{4\Delta^2}{g\omega_Q^2}$ is the inertia of the lattice mode [7, 8] and we have also added its damping rate $\gamma'$. This inertia term can be inferred from the microscopic CDW parameters via the gap equation $\Delta = 2\Lambda e^{-1/\lambda}$, where $\Lambda$ is the UV cutoff that is roughly the electronic bandwidth and $\lambda = g\nu_0$ is the dimensionless electron-phonon coupling constant. From experimentally extracted $\lambda$, $\Delta$ and $\omega_Q$ in representative quasi-one-dimensional CDW materials such as $K_{0.3}MoO_3$ [9] and TTF-TCNQ [10, 11], one has $\nu_{\rm ion} = \frac{4\Delta^2}{g\omega_Q^2} = 4\nu_0\frac{\Delta^2}{\lambda\omega_Q^2} \gg \nu_0$.

The kernel $\chi_{\sigma_3\sigma_3}(q,\omega)$ is the same as that for the EI phase mode up to a constant coefficient. We will only need its long-wavelength limit ($q \to 0$) that reads [12]

$$\chi_{\sigma_3\sigma_3}(\omega) = \sum_k \frac{1}{E_k}\frac{4\Delta^2}{\omega^2 - 4E_k^2} \simeq \nu_0 \begin{cases} -\dfrac{2\Delta}{\omega}\dfrac{\sin^{-1}(\frac{\omega}{2\Delta})}{\sqrt{1-(\frac{\omega}{2\Delta})^2}}, & \omega \le 2\Delta, \\ \dfrac{2\Delta}{\omega}\dfrac{\sinh^{-1}\sqrt{-1+(\frac{\omega}{2\Delta})^2}}{\sqrt{-1+(\frac{\omega}{2\Delta})^2}} - i\dfrac{2\Delta}{\omega}\dfrac{\pi}{2\sqrt{-1+(\frac{\omega}{2\Delta})^2}}, & \omega > 2\Delta. \end{cases} \tag{S42}$$

Here $\nu_0 = 1/(\pi v_F)$ is the normal-state density of states at the Fermi level. In the low frequency limit, one has $\chi_{\sigma_3\sigma_3} = -\nu_0$. From Eq. (S41), one obtains the dispersion of the phase mode which is linear in the low-energy regime: $\omega_q = \sqrt{n/[m(\nu_0 + \nu_{\rm ion})]}q$. Note that coupling to light converts the phase mode into a plasmon mode, which has square-root dispersion $\omega_q \propto \sqrt{q}$ in 2D and gapped dispersion in 3D [7]. However, these effects are small because of the large ion inertia $\nu_{\rm ion}$.

From Eq. (S41), one observes that there is linear coupling between phase $\theta$ and light already in the long wavelength limit. The equation of motion for the phase mode follows from the saddle-point condition $\delta S_{\rm eff}/\delta\theta^{\rm q} = 0$, yielding

$$-i\omega\theta_{\rm cl}(\omega) = \frac{\chi_{\sigma_3\sigma_3}(\omega)}{\nu_{\rm ion}(1 + i\gamma'/\omega) - \chi_{\sigma_3\sigma_3}(\omega)} v_F A(\omega) \equiv (\alpha_\omega - 1)v_F A\,. \tag{S43}$$

From it one obtains the dressed field

$$\tilde{A}_{\rm cl} = v_F A - i\omega\theta_{\rm cl} = \frac{\nu_{\rm ion}(1 + i\gamma'/\omega)}{\nu_{\rm ion}(1 + i\gamma'/\omega) - \chi_{\sigma_3\sigma_3}(\omega)} v_F A \equiv \alpha_\omega v_F A. \tag{S44}$$

Here $\alpha_\omega$ has the meaning of screening coefficient of the light field by the CDW phase mode. Because the CDW phase is locked to lattice sliding rather than a purely electronic phase, the large lattice inertia suppresses phase motion and makes the screening incomplete ($|\alpha_\omega| > 0$). From the current $J = -\delta S/\delta A_{\rm q} = -ev_F\chi_{\sigma_3\sigma_3}\tilde{A}_{\rm cl} - \frac{ne^2}{m}A$, one obtains the linear optical conductivity of the system:

$$\sigma(\omega) = \left(e^2 v_F^2 \alpha_\omega \chi_{\sigma_3\sigma_3} + \frac{ne^2}{m}\right)\frac{i}{\omega}\,. \tag{S45}$$

One observes that in the zero-ion-mass limit ($\nu_{\rm ion} = 0$), the first term is zero and the system exhibits 'superfluid' response with the optical weight being $ne^2/m$. This is because without the ion inertia, the electrons are interacting without retardation so that the whole system has Galilean invariance. Although the CDW order renders a quasiparticle gap, the whole system still behaves as an electron fluid whose center of mass responds to the uniform electric field as a free particle. Note that Galilean invariance leads to an even stronger statement: there are no nonlinear responses to a spatially uniform field. Nonzero ion mass makes the first term

nonzero, which shifts part of the optical weight to interband transitions at $\omega > 2\Delta$. In the low frequency regime $\omega \ll 2\Delta$, one has $\chi_{\sigma_3\sigma_3} = -\nu_0$ and the optical conductivity simplifies to $\sigma(\omega) = (1-\alpha)\frac{ne^2}{m}\frac{i}{\omega} = \frac{\nu_0}{\nu_{\rm ion}+\nu_0}\frac{ne^2}{m}\frac{i}{\omega+i\gamma}$ where $\gamma = \gamma'\nu_{\rm ion}/(\nu_{\rm ion}+\nu_0)$. Therefore, the low-frequency response is a Drude response with the remaining Drude weight $\frac{\nu_0}{\nu_{\rm ion}+\nu_0}\frac{ne^2}{m}$. In the infinite-ion-mass limit ($\nu_{\rm ion} = \infty$), one has $\alpha = 1$ and almost all of the Drude weight is transferred to interband transitions. Realistic materials have $\nu_{\rm ion} \gg \nu_0$. A typical spectrum of $\mathrm{Re}[\sigma(\omega)]$ is very similar to the $I_{\rm CDW}$ in Fig. 3(b) of the main text.

### B. The dynamics of the phase in the slow limit, effective force and photocurrents

We now construct the effective dynamics of the phase in the slow limit, meaning that the typical frequency is much smaller than the optical gap and the light frequency. Because there is high frequency linear response of the phase $\theta_{\rm cl}(\omega)$ to the optical field $A(\omega)$, this oscillation needs to be integrated out. Keeping only the slow and zero-wave-vector components of $\theta$ and $A_{\rm q}$, the effective Lagrangian is derived as

$$L_{\rm eff}[\theta_{\rm q/cl}; A_{\rm q}] = \theta_{\rm q}\left[\nu^*\left(\partial_t^2 + \gamma\partial_t\right)\theta_{\rm cl} - F_\theta\right] + A_{\rm q}\left[-v_{\rm F}\nu_0\partial_t\theta_{\rm cl} + I_{\rm inj} + I_{\rm drag}\right] \tag{S46}$$

where $\nu^* = \nu_0 + \nu_{\rm ion}$ is the net inertia of the phase mode.

$F_\theta$ is the light-induced NCSF on the uniform phase. Its leading order result is $F_\theta \propto q|E_\omega|^2$, which is at second order in the propagating optical field $E_\omega = -i\omega A_\omega$ and linear order in its wavevector $q$. From the original Lagrangian (Eq. (S38) and Eq. (S39)), this force has to arise from the $\partial_t\theta\sigma_3$ operator. The other terms involving $\theta$ all have spatial gradients, which would give zero for the uniform force on the phase. Because the force vertex is $\partial_t\theta\sigma_3$, the static force in the periodically driven case must come from the response kernel that diverges as $\propto 1/\omega'$ in the static limit ($\omega' \to 0$), which rules out all the bubble diagrams. Therefore, $F_\theta$ is contained only in the triangular diagram in Fig. S2(a) (Eq. (S12)), with the $\Gamma_{1/2}$ vertex given by Eq. (S39) and $\Gamma_3$ vertex (force vertex) being $-i\omega'\sigma_3$. Fortunately, only a subset of the terms from Eq. (S39) will contribute. First of all, one just needs the component of the triangular diagram $\chi_{\Gamma_1\Gamma_2\Gamma_3}(\omega, q; -\omega+\omega', -q; \omega')$ that diverges as $\propto 1/\omega'$ in the static limit. In the basis that diagonalizes the mean-field quasiparticle bands, this means that the matrix element from the force vertex must be an intraband one: the term $G_0(\Omega)G_0(\Omega+\omega')$ from Eq. (S12) would give $1/\omega'$ frequency integral only if they are from the same band. Secondly, for the $\Gamma_{1/2}$ vertex, one just needs to keep $\Gamma_{1/2} = \left(\partial_t\theta + v_{\rm F}A + \frac{1}{m}(\partial_x\theta)p\right)\sigma_3 + \frac{Ap}{m}\sigma_0$. In the terms we drop, the $A^2$, $(\partial_x\theta)^2$ and $A\partial_x\theta$ terms do not contribute to the triangular diagram at $O(A^2)$. The $\partial_x\theta\sigma_0$ term does not contribute at $O(q)$: the $\partial_x\theta$ contains a $q$, and the $\sigma_0$ means the interband matrix element from this vertex is also $O(q)$.

The $I_{\rm inj}$ in Eq. (S46) is the quasiparticle injection current induced by the optical driving. It satisfies

$$\frac{dI_{\rm inj}}{dt} = eK(\omega)|E_0|^2. \tag{S47}$$

Similar to the $F_\theta$, the injection current coefficient $K(\omega) \propto q$ is given by the component of the triangular diagram $\chi_{\Gamma_1\Gamma_2\Gamma_3}(\omega, q; -\omega+\omega', -q; \omega')$ that diverges as $\propto 1/\omega'$ in the static limit ($\omega' \to 0$). It is given by the triangular diagram in Fig. S2(a) while the same arguments for $F_\theta$ simplify the three vertices for it: $\Gamma_{1/2} = \left(\partial_t\theta + v_{\rm F}A + \frac{1}{m}(\partial_x\theta)p\right)\sigma_3 + \frac{Ap}{m}\sigma_0$ and $\Gamma_3 = J_{\rm P} = v_{\rm F}\sigma_3 + \frac{p}{m}\sigma_0$.

Comparing the diagrams for the force $F_\theta$ and the injection current, one observes that their only difference is a $\sigma_0 p/m$ term in the output vertex ($\Gamma_3$ vertex). Noting that $\sigma_0 p$ simply counts the total momentum of quasiparticles, one obtains the remarkable relation:

$$F_\theta = -\frac{1}{v_{\rm F}}K(\omega)|E_0|^2 + \frac{q}{mv_{\rm F}}S_{\rm abs}(\omega) \tag{S48}$$

where $S_{\rm abs}$ is the interband excitation rate of quasiparticles due to the drive. Furthermore, if the friction $\gamma$ of the phase mode vanishes, the only dissipative part in $\alpha_\omega$ will come from $\chi_{\sigma_3\sigma_3}$, see Eq. (S44). In this limit, from the $S_{\rm abs}$ calculated below in Eq. (S52), one will see that the excitation rate is set by the optical conductivity as $S_{\rm abs}(\omega) = 2\mathrm{Re}[\sigma(\omega)]|E_\omega|^2/\omega$, and Eq. (S48) becomes $F_\theta = \frac{1}{v_{\rm F}}\left(\frac{q}{2m\omega}\mathrm{Re}[\sigma(\omega)] - K(\omega)\right)|E_0|^2$.

#### 1. *Quasiparticle injection current*

It is well known that the free-particle injection current is [13]

$$\begin{aligned}\frac{dI_{\rm inj}}{dt} &= \frac{ev_{\rm F}}{\omega^2}\lim_{\omega'\to 0} i\omega'\chi_{\Gamma_1\Gamma_2\Gamma_3}(\omega, q; -\omega+\omega', -q; \omega') + (q, \omega \to -q, -\omega) \\ &= 2\pi e\sum_k |M(\omega, q; k)|^2\left[v_+(k+\frac{q}{2}) - v_-(k-\frac{q}{2})\right]\delta\left[\omega + E_-(k-\frac{q}{2}) - E_+(k+\frac{q}{2})\right]\end{aligned} \tag{S49}$$

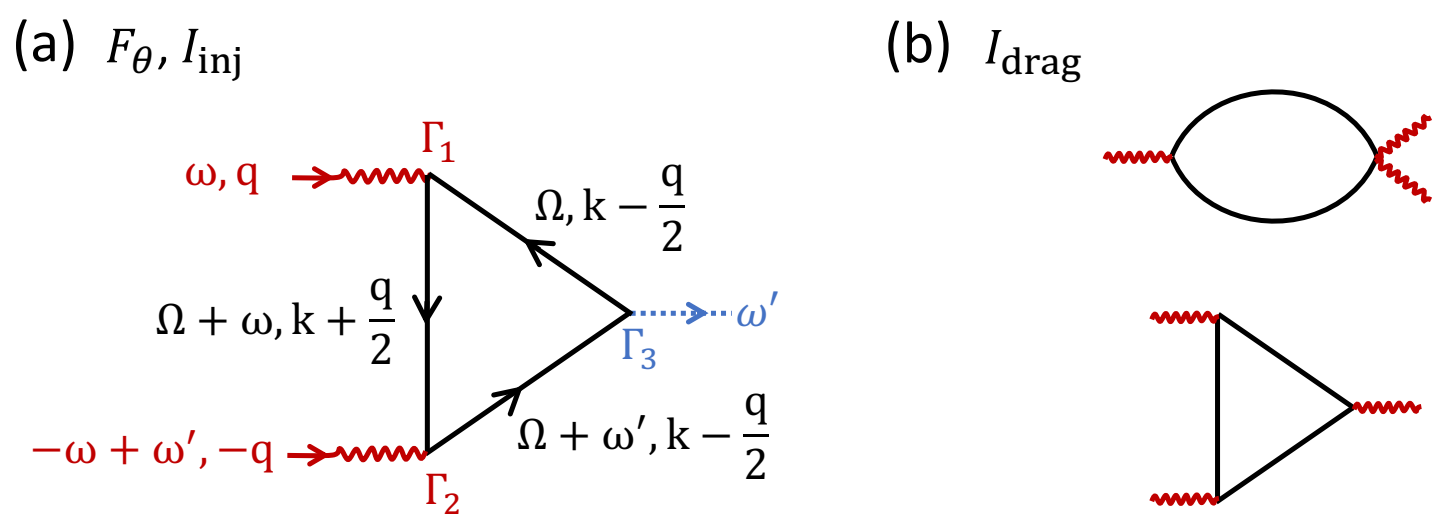


FIG. S2. (a) The triangular diagram that gives rise to the force on the CDW phase when the $\Gamma_3$ vertex is $-i\omega'\sigma_3$, or the quasiparticle injection current when the $\Gamma_3$ vertex is $J_{\rm P} = v_{\rm F}\sigma_3 + \frac{p}{m}\sigma_0$. The wavy red line represents the optical field plus the high frequency phase mode oscillations in Eq. (S39) that are in linear response to the optical field. These high-frequency phase modes do contribute to the diagram, which is defined as one-particle-irreducible (1PI) with respect to the slow ($\omega' \ll \omega$) phase field only. (b) The diagrams that give rise to the regular drag current.

where $M(\omega, q; k)$ is the interband matrix element of the $\Gamma_{1/2}$ vertex. Note that for the CDW system, this matrix element is not just that from the current operator, but also includes that of the oscillating phase field. Eq. (S49) can be simply understood from the Fermi golden rule: each resonant optical excitation results in an electron in the conduction band with group velocity $v_+$ and a hole in the valence band with group velocity $v_-$. Obviously, the injection current is zero if $q = 0$. Expanded to linear order in $q$, the matrix element reads

$$M(\omega, q; k) = -\frac{\Delta}{E_k}\left[\alpha_\omega + q\frac{k}{m}\left(\frac{1}{2E_k} - \frac{\alpha_\omega - 1}{\omega}\right) + O(q^2)\right] v_{\rm F} e A \tag{S50}$$

where $\alpha_\omega$ is defined in Eq. (S44). The change of group velocities of the quasiparticles is expanded to linear order in $q$ as $v_+(k + \frac{q}{2}) - v_-(k - \frac{q}{2}) = 2v_k + q/m + O(q^2)$ where $v_k = \partial_k E_k = v_{\rm F}^2 k/E_k$. The interband transition energy is expanded as $E_+(k + \frac{q}{2}) - E_-(k - \frac{q}{2}) = 2E_k + kq/m + O(q^2)$. Finally, the injection coefficient is obtained to linear order in $q$ as

$$K(\omega) = 8e^2 \mathrm{Re}[\alpha_\omega]\frac{q}{m}\frac{v_{\rm F}\Delta^2 |v_{\rm F} k_\omega|}{\omega^5}\Theta(\omega - 2\Delta) = 4\mathrm{Re}[\alpha_\omega]\frac{q e^2 v_{\rm F}}{m}\frac{\Delta^2\sqrt{\omega^2 - 4\Delta^2}}{\omega^5}\Theta(\omega - 2\Delta)\,. \tag{S51}$$

Here $k_\omega$ means the momentum at resonance. Remarkably, the terms proportional to $|\alpha_\omega|^2$ drop out as a result of cancellation. Note that the quasiparticle injection current is zero in the absence of the quadratic dispersion term (i.e., if $\xi_{k+k_F} = v_{\rm F} k$). In this limit, the quasiparticle band is particle-hole symmetric under $E_{+,k} \leftrightarrow E_{-,k}$ so that the theory and all phenomena should remain invariant upon changing the sign of the electron charge $e$. As the photocurrent is an effect proportional to $e^3$, it has to vanish. Indeed, one may check that for every interband transition from $(k_1, E_-)$ to $(k_2, E_+)$ that contributes a current, there is a transition from $(-k_2, -E_+)$ to $(-k_1, -E_-)$ that exactly cancels the former.

### 2. *Force on the phase*

The interband excitation rate of quasiparticles is found from the Fermi golden rule as

$$S_{\rm abs} = 2\pi\sum_k |M(\omega; k)|^2\delta(\omega - 2E_k) = |\alpha_\omega|^2\frac{\Delta^2}{\omega}\frac{2e^2}{|k_\omega|}\Theta(\omega - 2\Delta)|A_\omega|^2 = |\alpha_\omega|^2\frac{e^2 v_{\rm F}\Delta^2}{\omega^3\sqrt{\omega^2 - 4\Delta^2}}\Theta(\omega - 2\Delta)|E_0|^2\,. \tag{S52}$$

This expression may also be derived from the $O(1/\omega')$ part of the triangular diagram $\chi_{\Gamma_1\Gamma_2\Gamma_3}(\omega, q; -\omega + \omega', -q; \omega')$ with the $\Gamma_3$ vertex being $\sigma_0$.

The drag force on the phase is thus found from Eq. (S48), Eq. (S51) and Eq. (S52) as

$$\begin{aligned} F_\theta &= -\frac{1}{v_{\rm F}}4\mathrm{Re}[\alpha_\omega]\frac{q e^2 v_{\rm F}}{m}\frac{\Delta^2\sqrt{\omega^2 - 4\Delta^2}}{\omega^5}\Theta(\omega - 2\Delta)|E_0|^2 + \frac{q}{m v_{\rm F}}|\alpha_\omega|^2\frac{e^2 v_{\rm F}\Delta^2}{\omega^3\sqrt{\omega^2 - 4\Delta^2}}\Theta(\omega - 2\Delta)|E_0|^2 \\ &= \frac{q e^2 |E_0|^2}{m\omega^2}\frac{\Delta^2}{\omega\sqrt{\omega^2 - 4\Delta^2}}\left[-4\mathrm{Re}[\alpha_\omega]\frac{\omega^2 - 4\Delta^2}{\omega^2} + |\alpha_\omega|^2\right]\Theta(\omega - 2\Delta). \end{aligned} \tag{S53}$$

This force vanishes for the optical field below the optical gap. Above threshold, $\omega > 2\Delta$, the force becomes nonzero because of interband transitions.

The low-energy dynamics of the phase is thus $\nu^*(\partial_t^2 + \gamma\partial_t)\theta_{\rm cl} = F_\theta$ where $\gamma = \gamma'\nu_{\rm ion}/\nu^* \approx \gamma'$. Since the ionic effective mass is much larger than the electronic one, the phase dynamics is dominated by the ionic inertia. Balancing this force against the damping term in the steady state gives a dc CDW current:

$$I_{\rm CDW} = ev_{\rm F}\nu_0\partial_t\theta_{\rm cl} = \frac{e}{\pi}\partial_t\theta_{\rm cl} = \frac{e}{\pi}\frac{F_\theta}{\nu^*\gamma}\,. \tag{S54}$$

Because of the $1/\sqrt{\omega^2 - 4\Delta^2}$ factor, there is a sharp (divergent) onset of $I_{\rm CDW}$ once the light frequency is above the gap. This is rooted in the divergent density of states of quasiparticles just above the gap, making the optical absorption divergent. The frequency dependence of $\alpha_\omega$ regularizes the divergence in a way that is barely visible because of the large ion inertia $\nu_{\rm ion}$.

### 3. *Regular photon-drag current*

Among the second-order currents in Eq. (S46), in addition to the topological current $I_{\rm CDW} = ev_{\rm F}\nu_0\partial_t\theta_{\rm cl}$ and injection current $I_{\rm inj}$, there is also a regular photon-drag current $I_{\rm drag}$ shown as the last term in Eq. (S46). This photocurrent does not diverge as $1/\omega'$ as $\omega'$ goes to zero, and thus does not rely on a scattering rate to be finite. It has several contributions, which can be collected into the bubble diagram and triangular diagram (its regular component which does not diverge as $1/\omega'$ as $\omega'$ goes to zero) in Fig. S2(b).

From Eq. (S39), the electric current is $J = -\partial_A L = -\Psi'^\dagger\left(J_{\rm P} + J_{\rm D} + \frac{1}{m}\partial_x\theta\sigma_3\right)\Psi'$ where $J_{\rm P} = v_{\rm F}\sigma_3 + \frac{p}{m}\sigma_0$ is the paramagnetic current and $J_{\rm D} = \frac{n}{m}A$ is the diamagnetic current with $n$ being the normal state carrier density. The nonlinear current operator $(\frac{1}{m}\partial_x\theta\sigma_3)$ contributes additional cubic terms to the effective action:

$$S_{\rm eff} = \sum_{\omega,q}\left(\tilde{A}_q + \frac{1}{m}(A\partial_x\theta)_q\right)_{-\omega,-q}\chi_{\sigma_3\sigma_3}(\omega,q)\left(\tilde{A}_{\rm cl} + \frac{1}{m}(A\partial_x\theta)_{\rm cl}\right)_{\omega,q} \tag{S55}$$

after integrating out the fermions. Taking a derivative with respect to a static, uniform $A_{\rm q}$, one obtains an additional second order DC current:

$$\begin{aligned} I^{(1)}_{\rm drag} &= -\frac{1}{m}\tilde{A}_{\rm cl}(\omega,q)\chi_{\sigma_3\sigma_3}(\omega)\partial_x\theta_{\rm cl}(-\omega,-q) + v_{\rm F}\chi_{\sigma_3\sigma_3}(0)\frac{1}{m}A_{\rm cl}(\omega,q)\partial_x\theta_{\rm cl}(-\omega,-q) + (q,\omega\to -q,-\omega) \\ &= -\frac{q}{m\omega}v_F^2|A_\omega|^2\left[-\chi_{\sigma_3\sigma_3}(\omega)(|\alpha_\omega|^2 - \alpha_\omega) + \nu_0(\alpha_{-\omega} - 1) + (\omega\to-\omega)\right] \\ &= -2|A_\omega|^2\frac{q}{m\omega}v_F^2{\rm Re}\left[\chi_{\sigma_3\sigma_3}(\omega)(\alpha_\omega - |\alpha_\omega|^2) + \nu_0(\alpha_\omega - 1)\right]\,. \end{aligned} \tag{S56}$$

It corresponds to the bubble diagram in Fig. S2(b). In the low-energy limit ($\omega \ll 2\Delta$), this current has the expression $-\frac{1}{m}\nu_0\left(2v_{\rm F}A + \partial_t\theta\right)\partial_x\theta$, which may be understood as an emergent diamagnetic current with $\partial_x\theta$ being the local variation of charge density. Note that there are other contributions to the bubble diagram such as that from the diamagnetic current $J_{\rm D} = \frac{n}{m}A$. However, all these contributions are regular in the limit of $\omega' \to 0$. For light frequencies below the optical gap, $I_{\rm drag}$ is nonzero. The total photon-drag current of the CDW is therefore $I = I_{\rm CDW} + I_{\rm inj} + I_{\rm drag}$. In the Galilean-invariant limit (no phonon inertia so that $\alpha_\omega = 0$), the first two contributions vanish, and the drag current should reduce to $I_{\rm drag} = \frac{2nq}{m^2\omega}|A_\omega|^2$, which is that of a perfect electron fluid. Because $\alpha_\omega \approx 1$ in the realistic parameter regime, this regular current is small compared to the $I_{\rm CDW}$ and $I_{\rm inj}$ in a relatively clean system and we leave its detailed calculation to future research.


\fi

\end{document}